\documentclass[vecphys]{svmult}

\usepackage{makeidx}         
\usepackage{graphicx}        
\usepackage{multicol}        
\usepackage[bottom]{footmisc}
\usepackage{amssymb}
\usepackage{url}
\usepackage{cite}

\makeindex             

\begin{document}

\title*{Probing the Impact of Stellar Duplicity on Planet Occurrence with
Spectroscopic and Imaging Observations}

\author{Anne Eggenberger\inst{1,2} \and St\'ephane Udry\inst{1}}
\institute{Observatoire de Gen\`eve, Universit\'e de Gen\`eve, 51 ch. des Maillettes, 
CH-1290 Sauverny \and Laboratoire d'Astrophysique de Grenoble, 
Universit\'e Joseph Fourier, BP 53, F-38041 Grenoble Cedex 9 
\texttt{anne.eggenberger@obs.ujf-grenoble.fr,stephane.udry@obs.unige.ch}}
%
%
\titlerunning{Probing the Impact of Stellar Duplicity on Planet Occurrence}
\maketitle

\section{Introduction}

Over the last eleven years, Doppler spectroscopy has been very successful at 
detecting and characterizing extrasolar planets, providing us with a wealth of 
information on these distant worlds (e.g. \cite{Udry07a}). One important and considerably 
unexpected fact these new data have taught us is that diversity is the rule in 
the planetary world. Diversity is found not only in the characteristics
and orbital properties of the $\sim$210 planets detected thus far\footnote{See
the Extrasolar Planet Encyclopedia, \url{http://exoplanet.eu/}, for an 
up-to-date list.}, but also in the 
type of environment in which they reside and hence in which they are able to 
form. This observation has prompted a serious revision of the theories of 
planet formation (e.g. \cite{Durisen07,Lissauer07,Nagasawa07}), leading to 
the idea that planet formation may be a richer and more robust process than 
originally thought. 

It is well known that nearby G, K, and M dwarfs are more likely found in pairs 
or in multiple systems than in isolation. Specifically, 57\% of the G-dwarf 
primaries within 22~pc of the Sun have at least one stellar companion 
\cite{Duquennoy91}. The multiplicity among K dwarfs is very similar 
\cite{Halbwachs03,Eggenberger04b}, while the multiplicity among nearby M dwarfs 
is close to 30\% \cite{Fischer92,Delfosse04}. Altogether, these figures 
imply that more than half of the nearby F7--M4 dwarfs are in binaries or 
in higher order systems. Since these stars constitute the bulk of targets searched 
for extrasolar planets via Doppler spectroscopy, the question of the existence 
of planets in binaries and multiple stars is fundamental and cannot be avoided  
when one tries to assess the overall frequency of planets. 

From the theoretical perspective, the existence of planets in binaries and
multiple stars is not guaranteed a priori, as the presence of a stellar 
companion may disrupt both planet formation and long-term stability. 
On the other hand, young binary systems usually possess more than one 
protoplanetary disk, meaning that planets may form around any of the two 
stellar components 
(circumstellar planets) and/or around the pair as a whole (circumbinary 
planets). Although theoretically both circumstellar and circumbinary planets
should exist, our present observing programs are aimed at detecting circumstellar
planets and only these latter will be considered in this chapter. Our discussion 
will furthermore be focused on giant planets, which are less challenging to
detect by means of the Doppler spectroscopy technique than lower mass planets.

Two different scenarios have been proposed to explain the formation of gaseous 
giant planets. According to the core accretion model, giant planets form in a 
protoplanetary disk through the accretion of solid planetesimals 
followed by gas capture 
(see e.g. \cite{Lissauer07} for a review and references). Despite some remaining uncertainties, this model has 
the advantage of doing a fairly good job in explaining the existence of both 
the terrestrial and the giant planets in the Solar System, and so is  
considered as the favored formation 
mechanism for planets. With regard to planet formation in binaries, an 
important point in this model is that the protoplanetary cores that give rise 
to the giant planets have to form beyond the snow line (i.e. beyond 1--4~AU for 
solar-type stars) to benefit from the presence of ices as catalysts. 

An alternative way to view giant planet formation is to consider 
that gaseous giant planets form by direct fragmentation of the protoplanetary 
disk. This is the so-called disk instability model 
(see e.g. \cite{Durisen07} for a review and references). 
This scenario is still somewhat speculative in that it is not 
clear yet whether real protoplanetary disks actually meet the requirements to 
fragment. Furthermore, even if they do, it is far from 
certain that the fragments will be long-lived and contract into permanent 
planets. Given the many uncertainties and difficulties related to 
theoretical work on planet formation via disk instability, observational 
tests that would help characterizing and quantifying the likelihood of 
forming giant planets by this channel are highly desirable.

Regardless of the exact formation process, tidal perturbations from a close 
stellar companion may affect planet formation by truncating, stirring, and 
heating a potential circumstellar protoplanetary disk 
(e.g. \cite{Artymowicz94,Boss06,Mayer05,Nelson00,Pichardo05}). 
Disk truncation is a serious concern as it 
reduces the amount of material available for planet formation and as it may
cut the disk inside the snow line. This is a direct threat to planet formation  
and it explains why the naive outlook for planet formation in close 
binaries is pessimistic. The impact of disk stirring and heating on planet
formation is not so easily understood and requires dedicated simulations. 
Three main studies have been done so far, reaching somewhat
different conclusions as to the likelihood of forming giant planets in 
binaries closer than 50--60~AU. 
According to \cite{Nelson00}, giant planet formation is inhibited in 
equal-mass binaries with a separation of 50~AU whatever the formation 
mechanism, whereas \cite{Boss06} claims that giant planets are able to form 
in binaries with periastrons as small as 25~AU. On the other hand, 
\cite{Mayer05} showed that the protoplanetary disk mass has a strong impact on 
the final results and that the two possible formation 
mechanisms yield different predictions as to the occurrence of giant planets 
formed in light disks. This has a very interesting consequence, namely that 
planets in binaries might provide a unique data set to test theoretical 
predictions and to possibly identify the main formation mechanism for giant 
planets.

Assuming that planets can indeed form in various types of binary systems, 
another question is that of their survival. The
extensive body of literature on this subject can be summarized as follows.  
For low-inclination planetary orbits ($i$\,$\lesssim$\,$39^{\circ}$), the 
survival time is primarily determined by the binary periastron value and a 
stellar companion with a periastron wider than about 5--7 times the planetary 
semimajor axis does not constitute a serious threat to the long-term
($\sim$5 Gyr) stability of Jovian-mass planets 
(e.g. \cite{Fatuzzo06,Holman99}). The survival time of planets on higher
inclination orbits depends not only on the periastron value but also on the
inclination angle, meaning that planetary orbits become more easily unstable,
and this even if the periastron value is quite large (up to a few 
thousands of AU). This
additional type of instability is due to the so-called Kozai mechanism, which 
causes synchronous oscillations of the planet eccentricity and inclination 
(e.g. \cite{Innanen97,Takeda05}). 

To sum up, according to the present theoretical work if giant planets are to
form in binaries with a separation below $\sim$100~AU, then the most 
sensitive (but also less understood) issue regarding their occurrence in these 
systems is whether or not the planets can form in the first place. This
conclusion is very appealing, as it implies that quantifying the occurrence of
planets in close binaries may be a means of obtaining some observational
constraints on the process(es) underlying planet formation. However,
a word of caution is needed here. Recent work made to explain the existence of 
a close-in Jovian planet around HD\,188753\,A has emphasized the alternative 
possibility that close double and multiple star systems originally void of   
giant planets may acquire one via dynamical interactions (stellar encounters or 
exchanges), in which case the present orbital configuration of the 
system would not be indicative of the planetary formation process 
\cite{Pfahl05,PortegiesZwart05}. 
Pfahl \& Muterspaugh \cite{Pfahl06} have tried to quantify the likelihood that a binary system 
could acquire a giant planet in this way and concluded that 
dynamical processes could deposit Jovian planets in $\sim$0.1\% of the binaries 
closer than 50~AU. Therefore, to test the possibility of forming giant 
planets in binaries closer than $\sim$$50$~AU one needs not only to detect 
giant planets in these systems, but above all to quantify their occurrence.

From the observational perspective, the existence of planets in wide binaries 
and multiple stars has been supported by observations almost since the
first discoveries. Indeed, in 1997 three planets were found to orbit the 
primary components of wide binaries \cite{Butler97}, while another one 
was detected around 16\,Cyg\,b, the secondary component of a triple star 
system \cite{Cochran97}. The discovery of Gl\,86\,b a few years later 
\cite{Queloz00} was another milestone, as it showed that Jovian planets can 
also form and survive in the much closer spectroscopic binaries. This discovery 
prompted a new interest in the study of planets in binaries, raising the 
possibility that planets might be common in double and multiple star systems.

An important point to notice regarding the observation of planets in binaries
is that radial-velocity planet searches used to be, and still
are, strongly biased against the closest binaries. Double stars with an angular
separation smaller than a few times the size (projected onto the sky) of the 
spectrograph fiber or slit are indeed difficult targets for radial-velocity 
measurements, as the two components simultaneously contribute to the recorded 
flux. This not only introduces additional possibilities for spurious velocity 
variations, but it also makes it much more difficult to precisely extract the 
radial velocity of one component. Using standard
cross-correlation techniques, and even if only the radial velocity of the primary
star is of interest, Doppler searches for planets in binaries closer than 
$\sim$$2$--$6^{\prime\prime}$ become severely hampered, if not definitely ruled out. 
As a consequence, current data only provide sparse information on the account 
of the closest binaries as possible abodes for planets, and quantifying the 
frequency of planets in these systems remains impracticable.

Recognizing early the importance and the interest of including binary 
stars in extrasolar planet studies, we have been investigating the impact 
of stellar duplicity on planet occurrence for a few years. In this
chapter, we present some preliminary results from this dedicated 
investigation which has two main facets. One of our goals is to directly 
quantify the
occurrence of giant planets in binaries with very different separations, from
wide common proper motion pairs down to spectroscopic systems. Although close 
binaries are not well-suited targets for radial-velocity planet searches, 
dedicated reduction techniques based on two-dimensional correlation have
recently been developed to simultaneously extract the radial velocity of each
component \cite{Konacki05b,Zucker03}. In many 
instances, the precision achieved by these techniques is at least good enough to
search for Jovian planets around the primary star, meaning that Doppler surveys
for circumprimary giant planets in close binaries are feasible. By combining the
results from our ``classical'' radial-velocity planet searches conducted with
ELODIE \cite{Perrier03}, CORALIE \cite{Queloz00,Udry00} and HARPS \cite{Pepe04} 
with those from our dedicated survey for giant planets in single-lined
spectroscopic binaries \cite{Eggenberger03,Eggenberger07c} we should therefore 
be able to quantify the occurrence of giant planets in binaries as a function 
of the binary separation and to test some of the theoretical predictions
mentioned previously.

Another approach to the study of planets in binaries is to use direct imaging to
trace out how stellar duplicity impacts on planet occurrence
\cite{Eggenberger04c,Eggenberger07b,Eggenberger07c,Udry04}. 
For instance, if the presence of a close stellar companion hinders planet 
formation or drastically reduces the potential 
stability zones, then the frequency of planets in close binaries should  
be lower than the nominal frequency of planets orbiting single stars.
Alternatively, if the presence of a close stellar companion stimulates planet 
formation one way or another, planets should be more common in close binaries 
than around single stars. Reversing these statements, studying the multiplicity 
of planet-host stars relative to that of similar stars but without planetary 
companions may 
be a means of quantifying whether or not stellar duplicity has a negative 
impact on planet formation or evolution. 

This chapter is organized as follows. In Sect.~\ref{classical} we present the
results from classical radial-velocity planet searches, whose outcomes 
constitute the general framework within which lie more specific studies. In 
Sect.~\ref{imaging} we describe how direct imaging can be used to probe the global 
impact of stellar duplicity on planet occurrence and to test whether or not the 
frequency of planets is reduced in binaries closer than $\sim$120~AU. Finally, 
in Sect.~\ref{rvsearches_close} we discuss some preliminary results from our 
radial-velocity surveys dedicated to the search for circumstellar planets in 
spectroscopic binaries.


\section{Results from Classical Radial-Velocity Planet Searches}
\label{classical}

Since 1995, radial-velocity planet search programs have yielded a wealth of 
information about the properties of extrasolar planets, unveiling an 
outstanding variety of orbital parameters and characteristics that still 
challenges our views of planet formation (e.g. \cite{Marcy05b,Udry07a,Udry07b}). This observational material, in turn,
has been used quite extensively to get some insight into the formation and 
evolution processes at work in planetary systems 
(e.g. \cite{Eggenberger04,Halbwachs05,Santos03,Udry03}). Planets residing in
double and multiple star systems are particularly interesting targets in this
respect. Indeed, if the presence of a close stellar companion affects planet
formation or evolution as suggested by several theoretical studies, some imprints 
of these effects may be recorded in the properties and characteristics of the 
planets found in binaries and multiple stars. We show here that in spite of their 
discrimination against the closest binaries, classical radial-velocity 
planet searches have already provided us with important observational
constraints concerning the existence of planets in binaries.

\subsection{Selection Effects Against Close Binaries in Classical Doppler
Surveys}
\label{sel_effects_Doppler}

Doppler searches for planets around solar-type stars have always
avoided close binaries, though the meaning of the term ``close'' 
differs from one program to another \cite{Jones06,Marcy05,Perrier03,Udry00}. As for our ELODIE and CORALIE surveys, G 
and K dwarfs belonging either to ``short-period'' single-lined spectroscopic 
binaries ($\lesssim$10~years) or to double-lined spectroscopic binaries were systematically 
rejected from the main samples \cite{Perrier03,Udry00}. This discrimination 
was performed in the first place on the basis of former radial-velocity 
measurements 
gathered with the two CORAVEL\footnote{The two CORAVEL instruments 
\cite{Baranne79} were used extensively between 1977 and 1998 to monitor the 
radial velocity of more than 60\,000 nearby stars at an intermediate precision 
(typically 300~m\,s$^{-1}$) in both hemispheres.} instruments, but 
additional systems discovered later in the course of our 
planet surveys met the same fate and were rejected as well. 
Alternatively, single-lined spectroscopic binaries with long periods
($\gtrsim$10~years) were generally kept in the samples, since the presence of
giant planets is more likely in these systems.

Our initial policy on close visual binaries was less drastic and most of these  
systems were kept in our ELODIE and CORALIE samples. However, the data 
accumulated in the early phases of the CORALIE program showed 
that radial-velocity measurements of the primary components of close visual 
binaries were generally noisier and more variable than expected, suggesting that
the secondaries in these systems contribute to some extend to the recorded flux. 
Accordingly, visual binaries closer than $\sim$$6^{\prime\prime}$ and 
with a magnitude difference smaller than $\sim$4 were flagged as 
second-priority targets and observed less often than regular single stars.

\subsection{Census of Planets in Binaries and Multiple Star Systems}
\label{census}

Thanks mostly to recent searches for common proper motion companions to 
planet-host stars (Sect.~\ref{imaging}), the number of planets known to reside 
in binaries and multiple stars has been growing rapidely in the past few years 
\cite{Patience02,Eggenberger04,Mugrauer05a,Mugrauer06,Chauvin06,Raghavan06,
Desidera07,Eggenberger07b} and now
raises to 42 planets (or 35 planetary systems). A census of these planets 
is given in Table~\ref{pib_table}, which lists the planet-host systems with 
published planets and acknowledged evidence of a bound status for the stellar
components. 

\begin{table}[htb!]
\vspace{5mm}
\centering
\caption{Census of planets orbiting a component of a binary or multiple star 
system with confirmed orbital or common proper motion. CPM stands for common 
proper motion systems, VB for visual binaries showing hints of orbital motion,   
and SB for spectroscopic binaries. WD means that the stellar companion is a 
white dwarf.}
\scalebox{0.9}{
\setlength{\tabcolsep}{0.15cm}
\begin{tabular}{lcccl}
\hline
Planet-host star        & Proj. separation            & Number of   & Number of  & Remark \\
                       &  (AU)                        &  planets    &   stars   &       \\
\hline
HD\,38529\,A           & $\sim$12000       & 2 & 2  & CPM     \\
HD\,20782	       & $\sim$9113        & 1 & 2  & CPM     \\
HD\,40979              & $\sim$6400        & 1 & 2  & CPM     \\
HD\,222582             & $\sim$4740        & 1 & 2  & CPM      \\ 
HD\,147513             & $\sim$4441        & 1 & 2  & CPM, WD	 \\
HD\,213240\,A          & $\sim$3898        & 1 & 2  & CPM	\\ 
Gl\,777\,A             & $\sim$3000        & 2 & 2  & CPM	\\ 
HD\,89744\,A           & $\sim$2456        & 1 & 2  & CPM	\\ 
HD\,80606              & $\sim$1200        & 1 & 2  & CPM	\\ 
55\,Cnc\,A             & $\sim$1065        & 4 & 2  & CPM	 \\
HD\,11964\,A           & $\sim$1009        & 1 & 2  & CPM	 \\
16\,Cyg\,B             & $\sim$850         & 1 & 3  & CPM	\\ 
HD\,142022\,A          & $\sim$820         & 1 & 2  & CPM	\\ 
$\upsilon$\,And\,A            & $\sim$750         & 3 & 2  & CPM	\\  
HD\,178911\,B          & $\sim$640         & 1 & 3  & CPM	\\ 
HD\,75289\,A           & $\sim$621         & 1 & 2  & CPM	\\ 
HD\,196050\,A          & $\sim$501         & 1 & 3  & CPM	\\ 
HD\,99492              & $\sim$500         & 1 & 2  & CPM	\\ 
HD\,109749\,A          & $\sim$493         & 1 & 2  & CPM	\\ 
HD\,46375\,A           & $\sim$345         & 1 & 2  & CPM	\\ 
HD\,114729\,A          & $\sim$282         & 1 & 2  & CPM	\\ 
HD\,65216\,A           & $\sim$255         & 1 & 3  & CPM	\\ 
HD\,27442\,A           & $\sim$240         & 1 & 2  & CPM, WD	\\ 
$\tau$\,Boo\,A            & $\sim$240         & 1 & 2  & VB	\\ 
HD\,16141\,A           & $\sim$223         & 1 & 2  & CPM	\\ 
HD\,189733\,A          & $\sim$216         & 1 & 2  & CPM	\\ 
HD\,195019\,A          & $\sim$150         & 1 & 2  & CPM	\\ 
HD\,114762\,A          & $\sim$130         & 1 & 2  & CPM       \\
HD\,142\,A             & $\sim$105         & 1 & 2  & CPM	\\ 
HD\,19994\,A           & $\sim$100         & 1 & 2  & VB	\\ 
HD\,177830\,A          & $\sim$97          & 1 & 2  & CPM	\\ 
HD\,1237\,A            & $\sim$68          & 1 & 2  & CPM	\\ 
HD\,41004\,A           & $\sim$23          & 1 & 2  & SB	\\ 
$\gamma$\,Cephei\,A    & $\sim$22          & 1 & 2  & SB	\\ 
Gl\,86\,A              & $\sim$20          & 1 & 2  & VB, SB, WD\\
\hline
\end{tabular}}
\label{pib_table}
\end{table}

Table~\ref{pib_table} shows that most of the planets presently known to 
reside in binaries or multiple stars were found in systems with a
separation larger than $\sim$100~AU. Although some theoretical models predict a
shortage of giant planets in binaries closer than 100--120~AU, current Doppler
surveys are too severely biased against these particular systems to allow for a
definite conclusion. In particular, the fact that most of the few planets
detected in binaries closer than 100~AU were found in systems with separations 
of about 20~AU likely reflects the selection effects just mentioned in
Sect.~\ref{sel_effects_Doppler}. For instance, both $\gamma$\,Cephei and Gl\,86 are 
long-period single-lined spectroscopic binaries with very fait secondaries 
($\Delta V$\,$\sim$\,$8.4$ and $\Delta V$\,$\gtrsim$\,8, respectively). 
The presence of two such systems in our list confirms that 
searching for planets in long-period spectroscopic binaries with very faint 
secondaries is feasible using cross-correlation techniques. 
Similarly, HD\,1237 and HD\,177830 are visual binaries with very faint secondaries 
($\Delta V$\,$\gtrsim$\,$6$ and $\Delta V$\,$\gtrsim$\,$6.7$) that have not  
revealed themselves directly in Doppler measurements so far. 
As to the planet found around HD\,41004\,A ($\Delta V$\,$=$\,$3.7$), 
it is an object that is beyond the detection capabilities of classical Doppler 
surveys, but that is within the reach of surveys dedicated to
the search for planets in close binaries (Sect.~\ref{rvsearches_close}). 

The lack of planets in binaries closer than $\sim$\,20~AU is potentially more
interesting as it might be real. According to the most optimistic theoretical 
models, the closest binaries susceptible of hosting giant planets have a periastron
distance of about 25~AU \cite{Boss06}. If this is correct, then most ``short-period'' spectroscopic binaries 
should be free from giant planets and the ``limit'' at $\sim$20~AU may have a true 
meaning. Nonetheless, the present observing material do not allow us to rule out
the alternative hypothesis that the lack of planetary detections in systems
closer than 20~AU actually reflects the discrimination against ``short-period''
spectroscopic binaries in classical Doppler surveys. On that basis, the 
question of the closest binaries susceptible of
hosting circumstellar giant planets remains open. All we can say at 
present is that giant planets were found in all types of binaries where 
we have looked for them. 

Classical radial-velocity planet search programs have brought observational 
evidence that even spectroscopic binaries can
host circumstellar giant planets, a fact that was not taken for granted
previously. Nevertheless, the information provided by classical Doppler 
surveys is incomplete with regard to the closest binaries. 
Due to this incompletness, we can
derive from the present census only a minimum value for the fraction of planets 
residing in double and multiple stars. This minimum fraction is 21\%. Deriving 
the actual frequency of planets in binaries and probing the 
occurrence of planets in the closest systems both call for the 
need of planet search programs capable of dealing with spectroscopic and close 
visual binaries. Two such programs are presently underway 
\cite{Konacki05b,Eggenberger03}, and we discuss in Sect.~\ref{rvsearches_close} 
our own survey for planets in spectroscopic binaries.

\subsection{Different Properties for Planets in Binaries?}
\label{properties}

\begin{figure}[htb!]
\centering
\resizebox{11.5cm}{!}{
\includegraphics{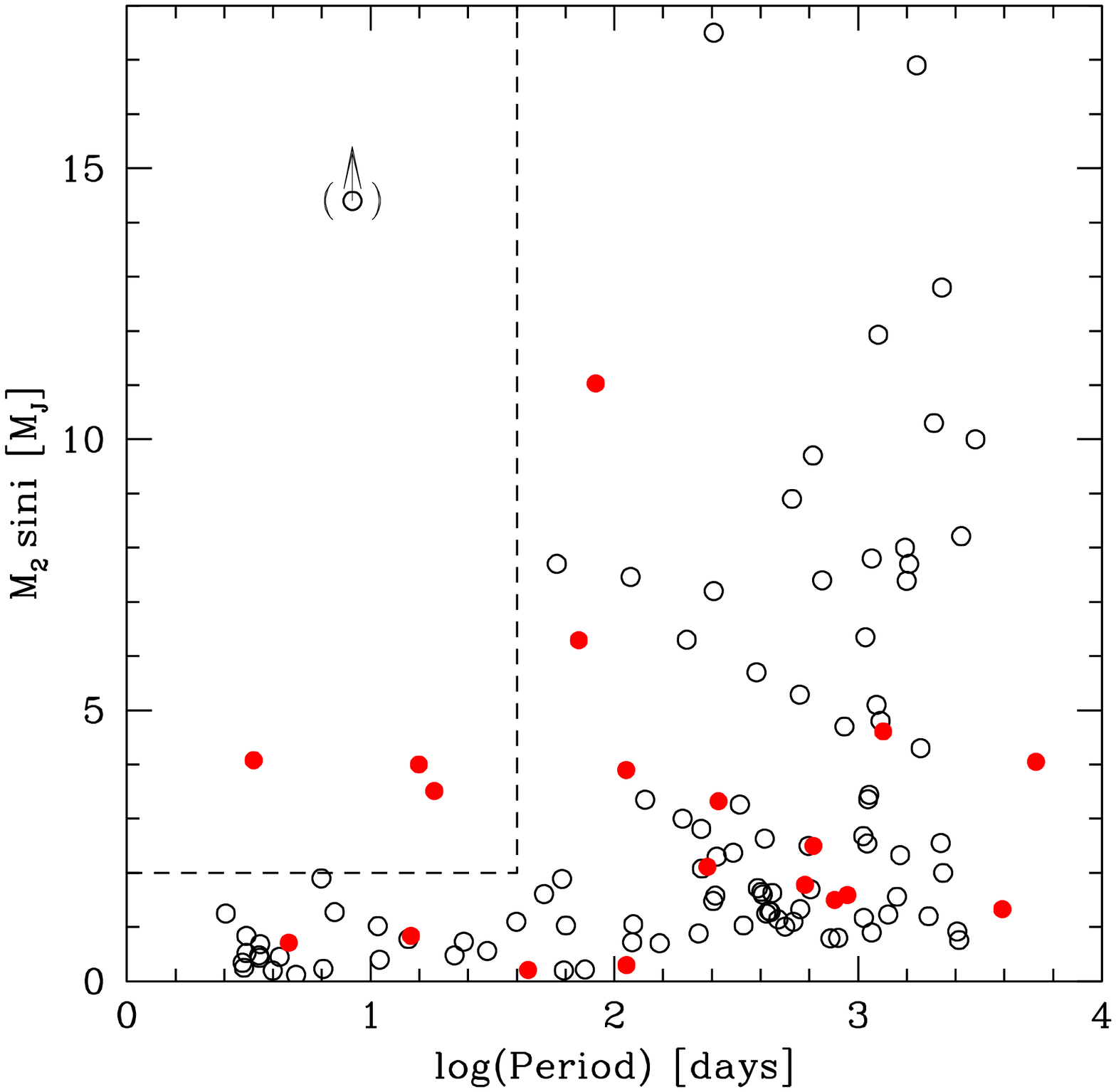}\hspace{0.1cm}
\includegraphics{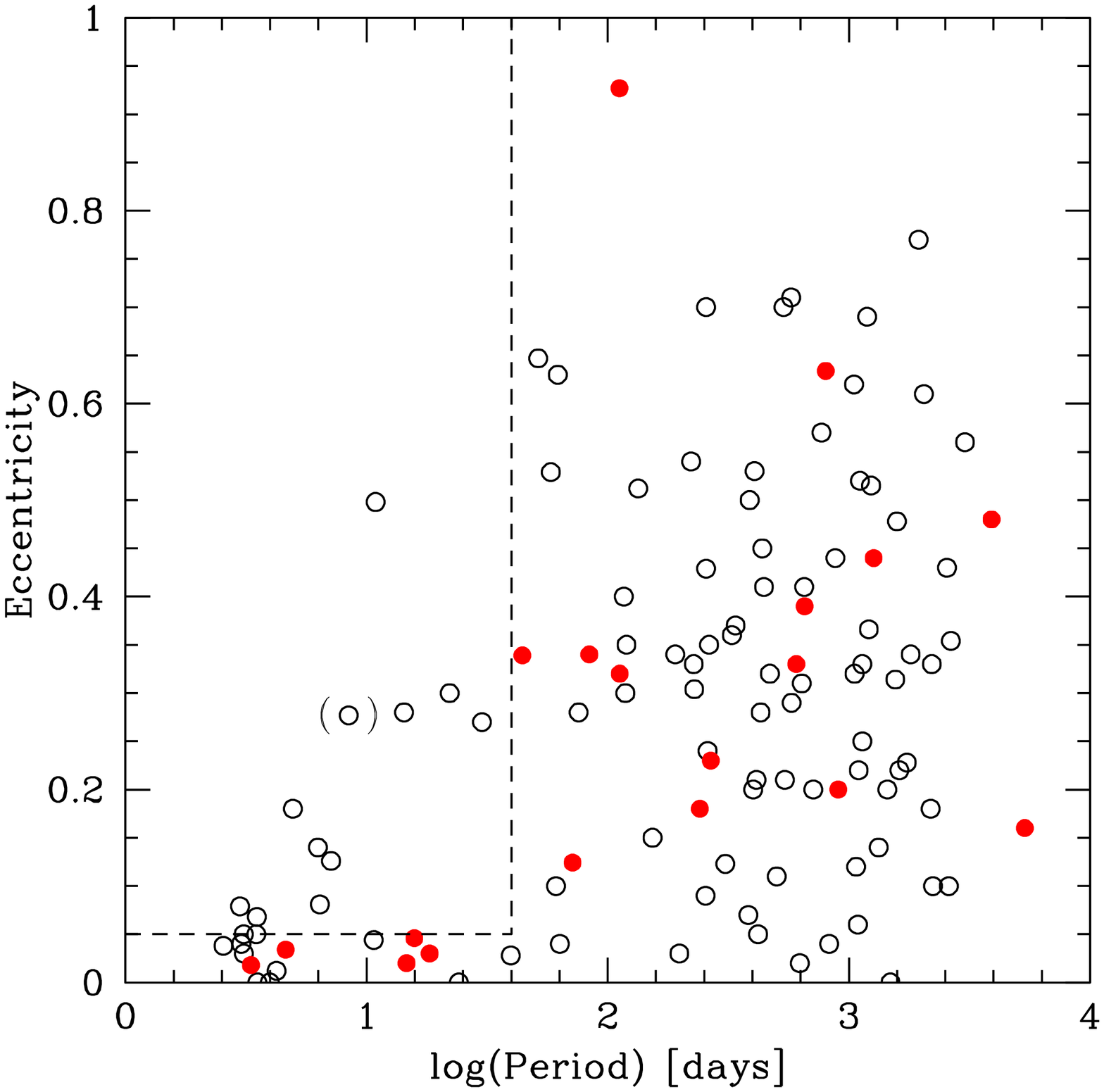}}
\caption{{\bf Left:} Minimum mass versus orbital period for all the extrasolar 
planetary candidates known in 2003. Planets orbiting a single star are 
represented as open circles, while planets residing in binaries or multiple 
star systems are represented as dots. The dashed line approximately
delimits the zone where only extrasolar planets belonging to binaries are found.
{\bf Right:} Eccentricity versus orbital period for the same planetary
candidates as before. The dashed line approximately delimits the region  
where no planet-in-binary is found.}
\label{stat3}
\end{figure}

The first hint that planets residing in binaries may possess some distinctive
properties and characteristics was brought by \cite{Zucker02}, who
pointed out that planets in binary systems seem to follow a different 
period-mass correlation than planets orbiting single stars. In a similar vein, 
we performed in 2003 a more comprehensive study based on a larger sample (19
instead of 9 planets in binaries), considering not only the period-mass 
but also the period-eccentricity diagram \cite{Eggenberger04}. As shown in 
Fig.~\ref{stat3}, our analysis confirms that the few planets with a minimum mass 
$M_2\sin{i}$\,$\gtrsim$\,$2$~M$_{\rm Jup}$ and a period 
$P$\,$\lesssim$\,$40$~days all orbit the components of binaries or multiple 
stars. However, the inclusion in our sample of several newly 
discovered planets with periods longer than 100~days, masses in the range 
3--5~M$_{\rm Jup}$, and found in binaries, decreases the significance of the 
negative period-mass correlation found by \cite{Zucker02}. Yet, 
marginal signs of this correlation subsist in the form of a shortage of 
very massive planets ($M_2\sin{i}$\,$\gtrsim$\,$5$~M$_{\rm Jup}$) on 
long-period orbits ($P$\,$\gtrsim$\,$100$~days). 

Regarding the period-eccentricity diagram, our analysis emphasizes that the 
planets with periods $P$\,$\lesssim$\,$40$~days and residing in binaries 
tend to have low eccentricities ($e$\,$\lesssim$\,$0.05$) compared to their
counterpart in orbit around single stars (Fig.~\ref{stat3}). 
In other words, the minimum period for a significant eccentricity seems larger 
for planets in binaries ($P$\,$\sim$\,40~days) than for planets around single 
stars ($P$\,$\sim$\,5~days). The statistical significance of this finding 
is very modest, though, and calls for confirmation.

Assuming that the above trends are real, is it possible to understand and to
explain these differences? 
 A study by \cite{Kley00} shows that the migration and 
mass growth rates of a Jovian protoplanet are enhanced when the latter is 
embedded in a circumprimary disk in a binary system with a mean
semimajor axis between 50 and 100~AU. At the same time, the eccentricity of such
a protoplanet decreases with time due to the damping action of the disk. 
Taken at face value, these theoretical predictions may 
provide a nice and self-consistent explanation for the observation 
that the most massive short-period planets are all found in binaries and have 
small eccentricities. However, the weak point in this reasoning is that the 
five planets with a period shorter than 40 days and found in binaries reside in 
systems with very different separations, from $\sim$20 to $\sim$1000\,AU. 
Although these separations are not
true semimajor axes (orbital parameters are unknown for most of theses 
binaries) and although it would probably be more appropriate to consider  
periastron distances rather than semimajor axes, it seems 
very likely that several of these five systems are quite different from the 
binaries modeled by \cite{Kley00}. 

An alternative explanation that would be valid for wide binaries is the 
so-called Kozai migration, a migration process specific to binaries and 
resulting from the coupling of the Kozai mechanism with tidal dissipation
\cite{Wu03}. Nonetheless, 
even if the Kozai mechanism can account for the high 
eccentricity of given planetary candidates such as 16\,Cyg\,Bb and HD\,80606\,b  
\cite{Holman97,Mazeh97,Wu03}, it has never been demonstrated that Kozai 
migration may account for the existence of close-in planets with very low 
eccentricities. In addition, several requirements must be simultaneously 
satisfied for the Kozai mechanism to operate, meaning that Kozai migration 
alone is unlikely to explain the distinctive characteristics of the five 
shortest period planets found in binaries. 

To summarize, the emerging trends seen in the period-mass and period-eccentricity
diagrams may be consistent with the idea that migration 
has played an important role in the history of short-period planets residing 
in binaries, and these results may be an indication that planetary migration proceeds 
differently in binaries than around single stars. However, owing to the very 
different types of systems they reside in, the properties of the five planets 
with a period shorter than about 40~days seem difficult to explain by invoking a 
single migration mechanism such as disk-planet interactions or Kozai migration.

As just shown, the first analyses devoted to the examination of 
possible differences in the characteristics and orbital properties of planets 
found in binaries and around single stars have come up with positive results. 
Yet, these analyses are far from being on robust statistical grounds and there 
is room for improvement (see \cite{Desidera07,Mugrauer05a} for more recent
analyses). One limiting factor in these studies is the small size 
of the planets-in-binaries sample that renders the results highly sensitive to 
the inclusion of additional new candidates. This observation, coupled to the
fact that short-period planets found in binaries may be more 
massive on average, constitutes an additional strong argument to promote the search for 
short-period planets in close binaries. 
Another weak point in these analyses is that they implicitly assumed that 
planet-host stars without known stellar companions were single stars, 
though the presence of stellar companions had never been systematically 
probed. Conducting systematic searches for stellar companions to the known 
planet-host stars constitute therefore another prerequisite step in more 
comprehensive studies aimed at investigating how companion stars affect 
planetary properties. This issue will be the subject of the next section.


\section{Results from Imaging Surveys}
\label{imaging}

As outlined in the introduction, the problem of quantifying the impact of 
stellar duplicity on planet occurrence can be tackled in a somewhat indirect 
way by comparing the multiplicity of planet-bearing stars to the multiplicity 
of similar stars but without known planetary companions. This approach was 
first followed by \cite{Patience02}, who probed the multiplicity status of
11 planet-host stars and concluded that the companion star fraction for
planet-bearing stars is not significantly different from that of field stars. 
Nonetheless, given the different outcomes and conclusions of theoretical studies on the 
formation of giant planets in binaries closer than $\sim$$100$~AU, and given 
that more than 170 planet-host stars are known today, the multiplicity of 
planet-bearing stars clearly merits reconsideration. 

To probe the multiplicity status of actual and potential planet-host stars, we 
initiated in 2002 a large-scale adaptive optics search for close stellar companions 
to nearby stars with and without known planetary companions 
\cite{Eggenberger04c,Udry04,Eggenberger07b,Eggenberger07c}. The main goal of
this program is to obtain a first quantification of the major effects of stellar
duplicity on planet formation and evolution, with an emphasis as to whether
or not the occurrence of giant planets is reduced in the presence of a close 
stellar companion. In order to access 
a large part of the celestial sphere, the main program was divided into two 
subprograms: a northern and a southern survey. The southern survey has been 
carried out with the NAOS-CONICA (NACO) facility (Very Large Telecope (VLT),
Paranal Observatory, Chile),
while the northern survey has been conducted with the PUEO-KIR adaptive optics 
system (Canada-France-Hawaii Telescope, Hawaii). In this section 
we present and discuss the results from our southern survey, which is the closest to 
completion.

\subsection{The Multiplicity Status of Nearby Stars With and Without Planets
Probed with VLT/NACO}
\label{naco_survey}

\subsubsection{The NACO survey}

In an effort to be as rigorous as possible, we included in our NACO survey both 
planet-host stars and comparison stars showing the least evidence for planetary 
companions from radial-velocity measurements. The inclusion of a comparison 
subsample was motivated by two main reasons. First, radial-velocity planet 
search programs suffer from noticeable selection effects against the closest
binaries and these biases must be taken into account to obtain meaningful 
results. Second, statistical studies must compare the multiplicity among
planet-host stars with the multiplicity among similar stars but without 
planetary companions. In addition, we needed a reference point spread function 
(PSF) star to characterize the adaptive optics system performance and to 
identify PSF artifacts on each of our images. To fulfill all these requirements 
at once, we selected a subsample of comparison stars within our CORALIE 
planet search sample and included these stars in our adaptive 
optics survey.
Proceeding in this way, we have at hand high-precision radial-velocity data 
that place constraints on the potential planet-bearing status of each 
comparison star, we match the target selection criteria for radial-velocity
planet searches, and we minimize the corrections related to observational 
effects.

Our NACO survey therefore relies on a sample of 57 planet-host stars,
together with 73 comparison stars carefully chosen so that they can be used both
as comparison stars for the scientific analysis and as PSF reference stars in
the data reduction process. Note that we purposely rejected from our 
observing list the 11 planet-host stars already observed by \cite{Patience02}. 
However, these stars will be included in our statistical analysis 
(Sect.~\ref{global_impact}), balancing 
the two subsample sizes to about 70 stars in each subsample.

The observing strategy of the survey consisted of taking a first image of each of our 
targets (planet-host and comparison stars) in order to detect companion
candidates. To discriminate between true companions and unrelated background or
forground stars, we 
relied on two-epoch astrometry. For relatively wide and bright objects 
($\rho$\,$\gtrsim$\,$10^{\prime\prime}$, $K$\,$\lesssim$\,$14$), a preexisting 
astrometric epoch could usually be found in the data from the Two Micron All 
Sky Survey (2MASS, \cite{Skrutskie06}), meaning that only one NACO observation 
was needed. Nevertheless, due to the high angular resolution and the small field of 
view of NACO, we could not rely on such preexisting data in a general way. 
As far as possible, our targets with companion candidates were thus observed 
twice during the survey to check for common proper motion.

\subsubsection{Detections and Observational Results}

Altogether, we found 95 companion candidates in the vicinity of 33 of our
targets. Among the 61 companion candidates with multiepoch observations, 19 are 
true companions and 1 is likely bound. The companionship status of the 34 
companion candidates with only one observing epoch remains formely unknown, but 
calculations of individual likelihood of chance alignment show that most of 
these objects are likely background stars (see \cite{Eggenberger07b} for further details).

Among planet-host stars, we discovered two new companions to HD\,65216, 
one to HD\,177830, and we resolved the previously known companion to 
HD\,196050 into a close pair of M dwarfs. Our data additionally confirm the 
bound nature of the companions to the planet-host stars HD\,142, HD\,16141 
and HD\,46375 (Table~\ref{pib_table}), along with the unbound status of a close
and relatively bright companion to HD\,162020. 
The two companions to HD\,65216 form a tight binary 
(projected separation of 5~AU) and are probably very low mass stars at the 
bottom of the main sequence. HD\,65216 therefore joins the small group of triple star
systems hosting a planet. The same is true for HD\,196050, which was previously
thought to be a binary \cite{Mugrauer05a}, but turned out to be a triple when 
observed at the NACO resolution. 
Like the triple system HD\,178911, HD\,65216 and HD\,196050 are made of a 
single planet-host star in orbit with a more distant binary, but contrary to 
HD\,178911 they contain stars with very different masses (G primaries,  
M/L secondaries and tertiaries). As the discovery a such systems requires 
high angular resolution and high-contrast data, it is not very surprising that 
we found two in our NACO survey whereas none was previously known (or 
recognized as such). 
The newly found companion to HD\,177830 is an early M dwarf 
located at a projected separation of 97~AU. Note that HD\,177830 is not 
a main-sequence star but an evolved K0 subgiant.

\subsubsection{Survey Sensitivity and Parameter Space Surveyed}

For G0 primaries, the typical sensitivity of our survey enabled us to 
detect companions down to M4--M5 dwarfs at $0.2^{\prime\prime}$ and all M dwarf 
companions above $0.65^{\prime\prime}$. For K0 primaries, we detected companions 
down to M5--M6 dwarfs at $0.2^{\prime\prime}$ and we reached the substellar domain 
above $0.65^{\prime\prime}$. Our survey thus provides us with a very
complete census of the stellar multiplicity among planet-host stars for mean 
semimajor axes in the range 35--230~AU, 
allowing us to probe a large fraction of the most interesting and 
sensitive separation range according to planet-formation models (i.e. the region 
below and around 100--120~AU).

\subsection{The Global Impact of Stellar Duplicity on Planet Occurrence}
\label{global_impact}

The observational results obtained in the context of our NACO survey form an 
unprecedented data set to study the impact of stellar duplicity on planet 
formation and evolution. Indeed, adding to our own results the targets surveyed 
by \cite{Patience02} we now have a precise and homogeneous 
census of the multiplicity status of 68 planet-host stars. 
As importantly, we now also have a comparison subsample of 73 stars with both 
high-precision radial-velocity measurements and adaptive optics data, so
that we can effectively compare some of the statistical properties of 
planet-bearing stars with those of similar stars showing no evidence for
planetary companions. This statistical analysis is currently in progress. We 
present here a preliminary and simplified version aimed at obtaining a first 
quantification of the global impact of stellar duplicity on planet occurrence 
in binaries with mean semimajor axes between 35 and 230~AU.

\begin{figure}[htb!]
\centering
\resizebox{11.5cm}{!}{
\includegraphics{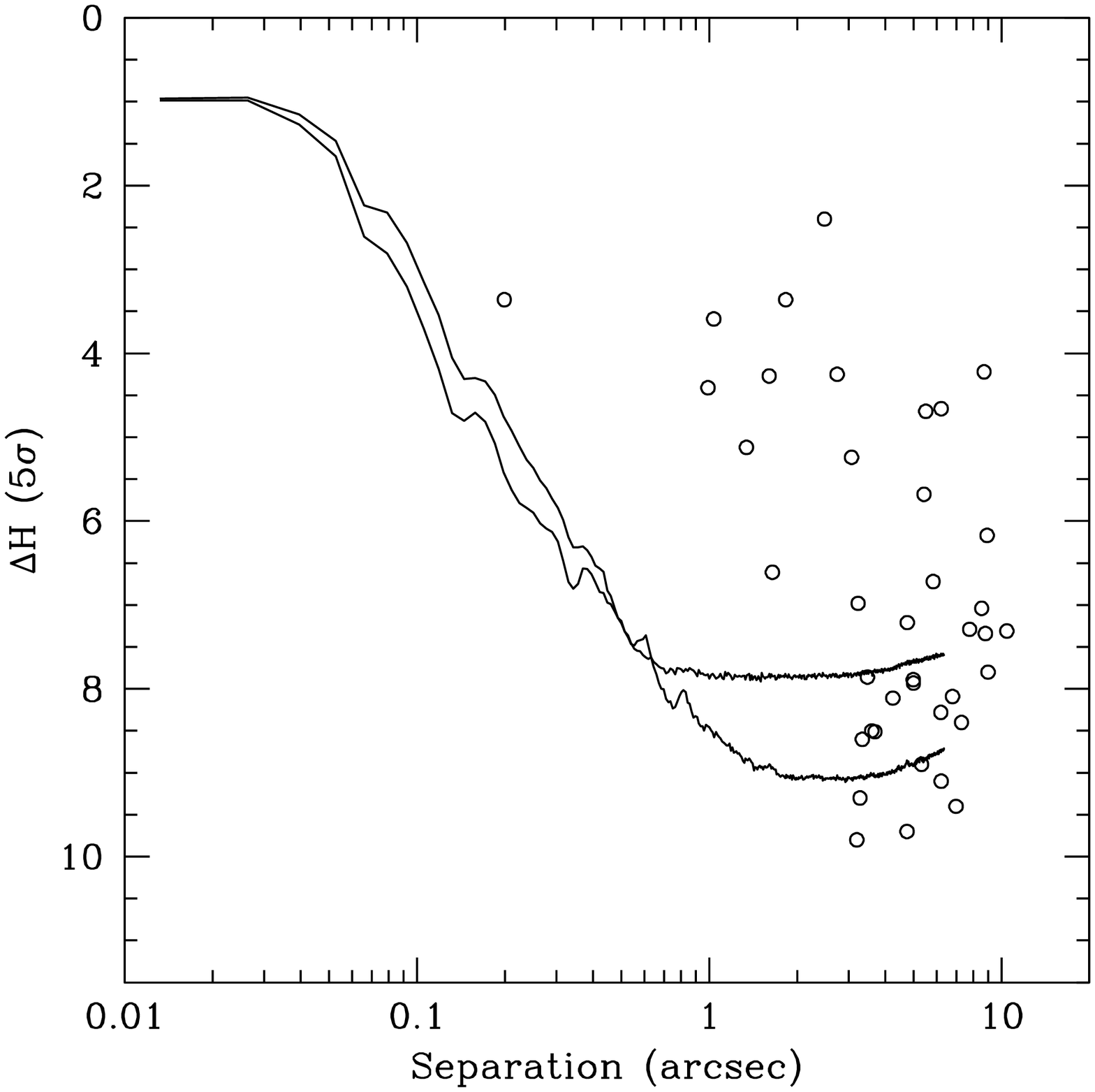}\hspace{0.1cm}
\includegraphics{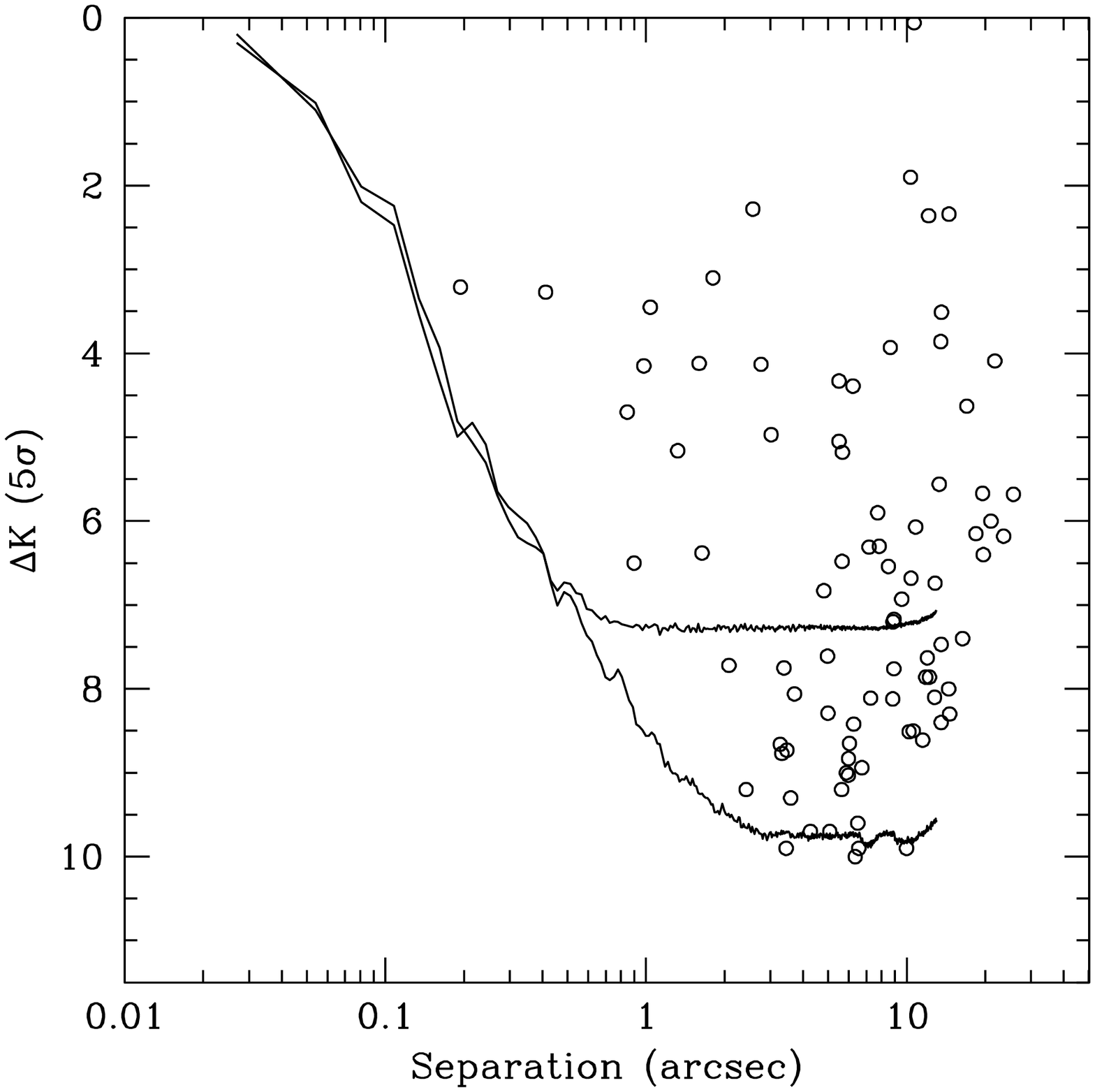}}
\caption{Summary of the detections (open circles) and median sensitivity limits 
(solid lines) from our NACO survey in the $H$ (left) and $K$ (right) bands. 
The original detector of CONICA was replaced in the middle of our program, hence
two sensitivity limits for each band.}
\label{detlim_survey}
\end{figure}

As already mentioned, Doppler planet searches are biased in some ways against
the closest binaries and these selection effects must be taken into account to
obtain meaningful results. 
The two main selection effects associated with Doppler planet search programs 
are visible in Fig.~\ref{detlim_survey}, which shows the detections and
sensitivity limits from our NACO survey. This figure emphasizes a 
shortage of systems with angular separations below $\sim$$0.8^{\prime\prime}$. 
This paucity of very close companions is neither real nor due to a bad 
estimation of our detection limits, but simply reflects the systematic 
rejection of most spectroscopic binaries from the CORALIE sample. 
Another striking feature in Fig.~\ref{detlim_survey} is the small number of 
companions with a magnitude difference smaller than 3 in the $H$ band and 
smaller than 2--3 in the $K$ band. This is a signature of the rejection of 
close visual binaries with bright secondaries, i.e. the systems with an 
angular separation below $\sim$$6^{\prime\prime}$ and a magnitude 
difference in the $V$ band smaller than $\sim$4, which translates into a 
magnitude difference of about 2.2 in the $K$ band. Since we were very careful 
with the selection of our comparison subsample, our two NACO subsamples are 
biased in the same way with respect to these selection effects and the results 
obtained for planet-host and for comparison stars can be compared 
directly, at least for a preliminary analysis.

In order to quantify the global impact of stellar duplicity on planet 
occurrence we use the binary fraction. As explained in the introduction, if 
stellar duplicity hampers planet formation or threatens the long-term survival 
of planets, the binary fraction of planet-host stars should be smaller than the
binary fraction of comparison stars. Alternatively, if the presence of a close
stellar companion favors planet formation one way or another, the binary
fraction of planet-host stars should be greater than that of comparison stars. 

As a first step in the analysis we computed the binary fraction for both 
subsamples, considering true and likely bound companions with angular 
separations in the range 0.8--6.5$^{\prime\prime}$ and located within the most 
restrictive detection limits of either of the two bands ($H$ and $K$) used in 
our survey. On that basis, the binary
fraction for planet-host stars is $8.8$\,$\pm$\,$3.5$\%, while the binary
fraction for comparison stars is $12.3$\,$\pm$\,$3.2$\%. The two binary 
fractions are thus basically compatible within error bars, though
the binary fraction for planet-host stars might be slightly smaller. In any case,
these results show that the presence of a stellar companion with a mean semimajor
axis between 35 and 230~AU does not favor planet occurrence to a significant
extent. 

To investigate in more detail the potentially negative impact of stellar 
duplicity on planet occurrence in the closest systems, we divided our 
two subsamples in two, following theoretical predictions that place a limit at
100--120~AU between the regime of close binaries in which planet formation might
be affected by stellar duplicity, and wider binaries in which planet formation
proceeds like around single stars. Recomputing the binary fractions for our two subsamples, but 
considering only the systems closer than 120~AU, yields a binary fraction 
of $2.9$\,$\pm$\,$2.1$\% for planet-host stars and of 
$9.6$\,$\pm$\,$3.5$\% for comparison stars. Similarly, the two binary fractions 
for the wider systems are $5.9$\,$\pm$\,$2.0$\% for planet-host stars and 
$2.7$\,$\pm$\,$2.9$\% for comparison stars. According to these results, the
binary fractions for planet-host and comparison stars are compatible for 
binaries with a mean semimajor axis between 120 and 230~AU, while they 
differ at the $1.6$$\sigma$ level for closer binaries. The main result of  
our analysis is thus that the occurrence of planets is reduced in binaries closer 
than $\sim$120~AU. Even though the statistical significance of this result is 
quite modest, this is an important finding.

Given the range of semimajor axes considered in our analysis, the lower frequency 
of planets in binaries closer than $\sim$120~AU is likely to be
related to the formation of the planets rather than their long-term 
survival. Recalling the conclusions from theoretical studies, one possible 
explanation to our findings would be that disk instability is indeed a viable
mechanism for the formation of giant planets and that this mechanism gets
inhibited in binaries closer than $\sim$120~AU, as suggested by \cite{Mayer05}. 
However, one weak point in this reasoning is that \cite{Mayer05} did not
actually study planet formation via core accretion. Their prediction that 
the formation of giant planets via core accretion proceeds undisturbed in 
binaries with
separations down to $\sim$60~AU is based solely on the temperature profiles of
their simulated disks, while additional effects may come into play to inhibit
planet formation. Consequently, our results might alternatively indicate that
core accretion is the main formation mechanism for giant planets, but that its
efficiency is reduced in binaries closer than $\sim$120~AU. This point of view 
may be consistent with the conclusion by \cite{Thebault04} that planetesimal 
accretion is possible in the $\gamma$\,Cephei system (semimajor axis of 18.5~AU), 
but requires a delicate balancing between gas drag and secular perturbations by 
the secondary star.

To sum up, the present results from our NACO survey constitute the first
observational evidence that the frequency of planets is reduced in binaries 
closer than $\sim$120~AU. Nonetheless, further investigations on both the 
theoretical and the observational sides will be needed to put this result on
robust statistical grounds and to fully explain the origin of this lower 
frequency. Regarding observations, we are working on a more comprehensive and 
refined version of the preliminary analysis just presented, also taking into 
account the recent results from the survey by \cite{Chauvin06}. The next step 
in the analysis will be to
include the results from our PUEO northern survey. On the one hand this will provide us
with an improved statistics to study the global impact of stellar duplicity on
planet occurrence, while on the other hand this new material will enable us to
reconsider the emerging trends outlined in Sect.~\ref{properties} regarding possible 
differences in the properties of short-period planets residing in binaries. Last
but not least, we would like to better characterize the implicit assumption 
made throughout our discussion that most planet-host stellar systems closer 
than $\sim$120~AU have retained their current orbital configuration ever since
the planets formed. As explained in the introduction, a direct quantification of 
the occurrence of giant planets in the closest binaries might provide some
observational constraints on this point. 


\section{Results from Radial-Velocity Planet Searches in Spectroscopic Binaries}
\label{rvsearches_close}

Planet searches in close binaries (i.e. systems closer than 
2--6$^{\prime\prime}$) used to be of marginal interest until
2000--2002. The discovery of Gl\,86\,b, the first planet found in a
spectroscopic binary \cite{Queloz00}, and the observation that the most 
massive short-period planets all orbit the components of double or multiple 
star systems \cite{Udry02,Zucker02} changed this
point of view and led to an ever-increased interest for planet searches in 
close binaries. Yet, classical Doppler surveys do not avoid most close 
binaries without reason. The main issue with close binaries is that 
each stellar component cannot be observed individually. That is, Doppler data of
close binaries consist not of single stellar spectra but of composite 
spectra made of two (or possibly more) stellar spectra. Obviously, this 
introduces some complications into the extraction of the radial
velocities, rendering classical cross-correlation techniques not well adapted to
the search for circumstellar planets in close binaries. The inclusion of close 
binaries into radial-velocity planet searches thus necessitated the development
of data reduction techniques specially designed to extract precise radial 
velocities from composite spectra.

A rather natural
way to analyze composite spectra and to extract precise radial velocities for 
the individual components of close binaries is to generalize the concept of 
cross-correlation to that of two-dimensional correlation. This approach was 
followed some time ago by S. Zucker and T. Mazeh, who developed a 
two-dimensional correlation algorithm named TODCOR \cite{Zucker94}. 
Following our intention to include spectroscopic binaries in our 
radial-velocity planet searches, we teamed up with S. Zucker and T. Mazeh, who 
modified their TODCOR algorithm to allow it to work with our ELODIE and 
CORALIE echelle spectra. This resulted in a new multi-order TODCOR algorithm
\cite{Zucker03}, which has already produced some very interesting results 
\cite{Zucker03,Zucker04,Eggenberger07a,Eggenberger07c} and 
which we are now using extensively to search for planets in spectroscopic and
close visual binaries. 

We present and discuss in this section some results from our ongoing searches
for planets in spectroscopic binaries. Our presentation will follow an
increasing order of difficulty in terms of radial-velocity extraction, starting
with the easiest systems that are single-lined spectroscopic binaries 
(SB1s) and ending with the more complicated double-lined spectroscopic 
binaries (SB2s).

\subsection{Searching for Planets in SB1s: Our Survey for Short-Period
Circumprimary Planets}
\label{sb1s}

In order to obtain a first quantification of the occurrence of planets in the
closest binaries susceptible of hosting circumstellar planets we initiated in
2001 a systematic radial-velocity search for short-period circumprimary  
planets in SB1s \cite{Eggenberger03,Eggenberger07c}. 
The restriction of our survey to SB1s was motivated by
two considerations. First, the faintness of the secondary components in these
systems gave us good hopes that we could use our standard cross-correlation 
technique to extract precise radial velocities for the primary components. 
Second, the prospects of planet formation and survival might be brighter 
in SB1s than in SB2s, which have similar separations but 
more massive secondaries. Our survey for 
giant planets in SB1s was thus designed as a first exploratory investigation 
that may be complemented later, in the case of positive results, by an 
additional survey targeting SB2s .

\subsubsection{Sample and Observations}

Our sample of binaries was selected on the basis of former CORAVEL surveys 
carried out to study the multiplicity among  
G and K dwarfs of the solar neighborhood \cite{Duquennoy91b,Halbwachs03}. 
Basically, we retained all the 140 SB1 candidates with a period longer 
than $\sim$$1.5$\,years, some of them with well-characterzied orbits,  
others with long-period drifts. Note that CORAVEL 
velocities have a typical precision of 300~m\,s$^{-1}$ and thus cannot be used 
to search for planets. To search for planets in our 140 SB1s we took 10 to 15 
additional high-precision radial-velocity measurements of each system, either 
with the ELODIE spectrograph (Observatoire de Haute-Provence, France;
\cite{Baranne96,Perrier03}) or with 
the CORALIE spectrograph (La Silla Observatory, Chile;
\cite{Queloz00,Udry00,Pepe02}). Given our initial aim 
to analyze these high-precision data with standard cross-correlation techniques, 
we rejected during the observations the systems that turned out to be SB2s at 
the higher resolution of ELODIE and CORALIE, as well as the binaries that were 
resolved within the guiding field of the telescope. After this additional 
selection we ended up with 101 SB1s that form the core of our survey.

\subsubsection{First Analysis Based on Cross-Correlation}

As a first step in the analysis, the spectra obtained with ELODIE and CORALIE 
were reduced online and the radial velocities extracted using our standard 
cross-correlation pipeline. When searching for planets in binaries, what we are 
interested in are not the radial velocities themselves but the residual 
(radial) velocities around the binary orbits. The planet search was thus 
carried out by searching for short-period variations in these residual velocities.

\begin{figure}[htb!]
\centering
\resizebox{7.0cm}{!}{
\includegraphics{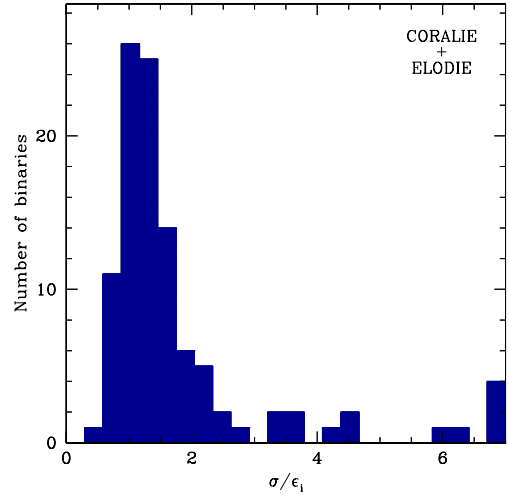}}
\caption{Normalized residual-velocity rms for all our SB1s. 
$\sigma$ is the standard deviation around a Keplerian orbit or around a drift,  
while $\epsilon$ is the mean measurement uncertainty. Systems with a rms larger
than 7 are all gathered together in the last bin.}
\label{pibsig}
\end{figure}

\begin{figure}[htb!]
\centering
\resizebox{11.5cm}{!}{
\includegraphics{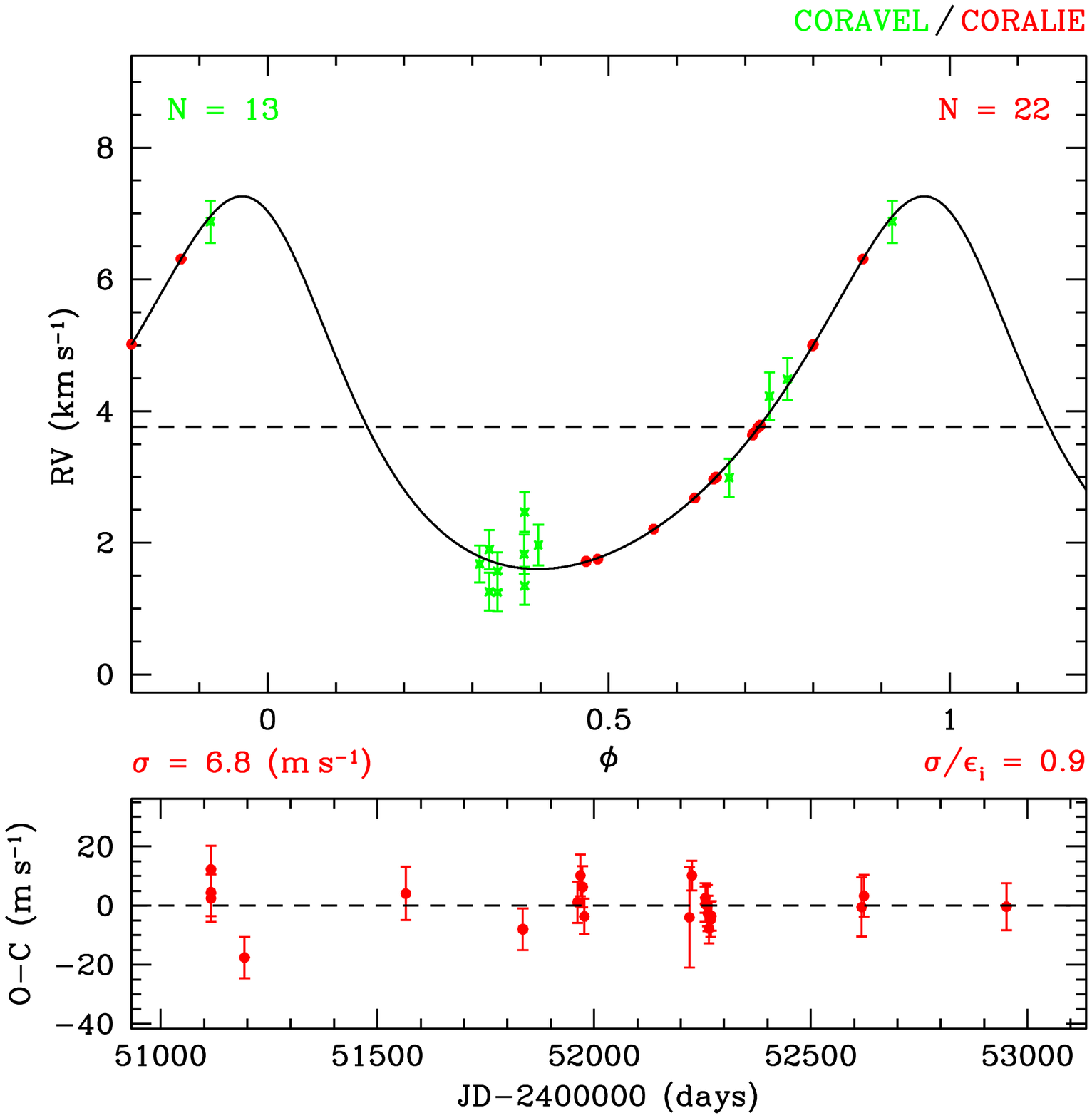}\hspace{0.1cm}
\includegraphics{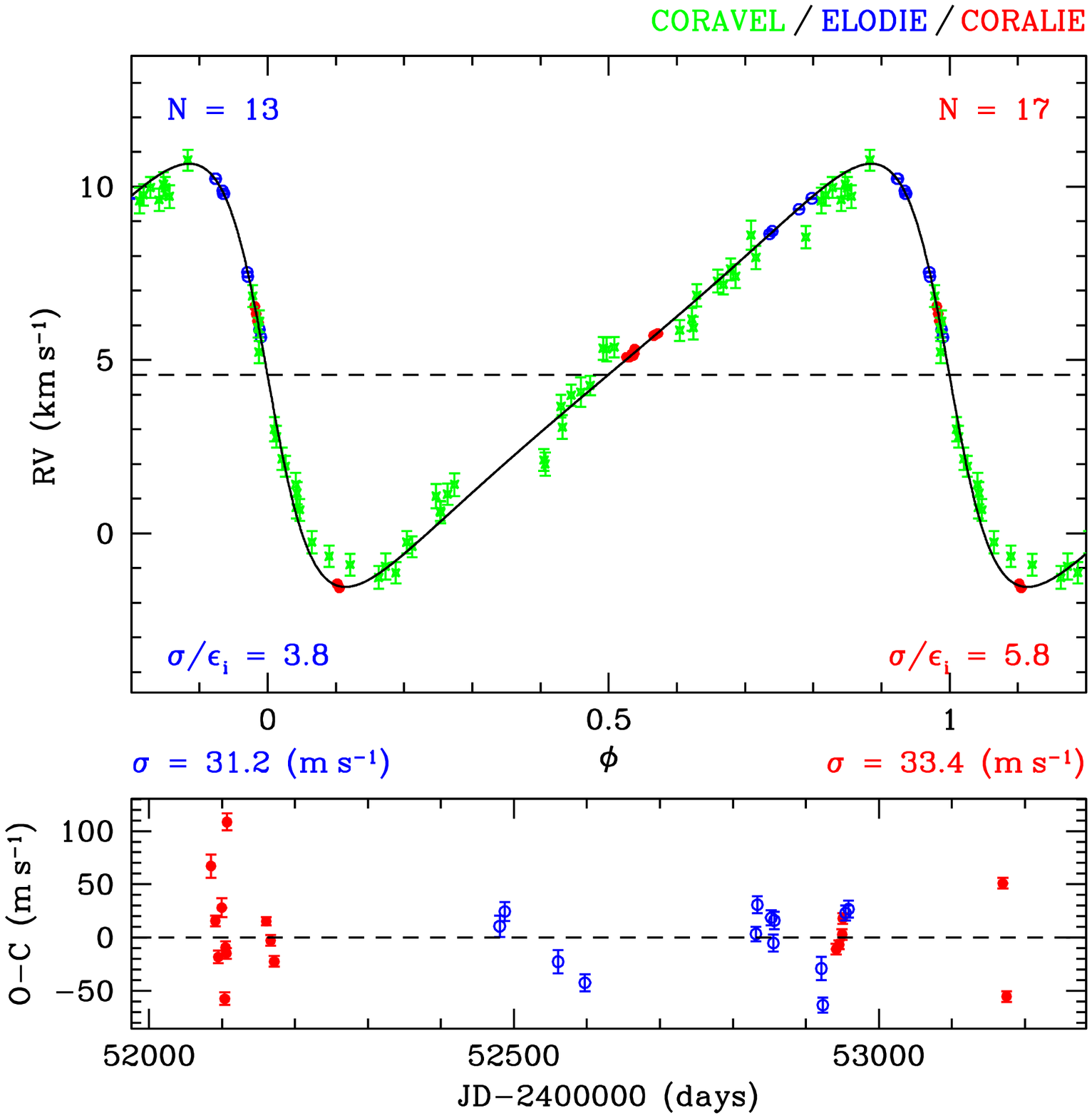}}
\caption{{\bf Left:} Example of a binary exhibiting no residual-velocity
variation. CORAVEL data are depicted as stars (large error bars), while CORALIE
data are depicted as dots. The bottom panel shows the residual velocities
(CORALIE data only). 
{\bf Right:} Example of a binary with variable residual velocities. 
This system was exceptionally observed with both ELODIE and CORALIE. 
Figures on the left refer to the ELODIE velocities (represented as circles), 
while figures on the right refer to the CORALIE velocities (represented as 
dots).}
\label{exs_pib}
\end{figure}

Figure~\ref{pibsig} shows the distribution of the residual-velocity variations 
for our 101 targets. These variations are quantified by a normalized
root-mean-square (rms), 
which is the ratio of the external error (i.e. the standard deviation around 
the orbit or around the drift) to the mean internal error (i.e. the mean of 
individual photon-noise errors). According to 
Fig.~\ref{pibsig}, most of our targets (74\%) have a normalized rms close to 
1, indicating that no source of radial-velocity variation other than the 
orbital motion is present (see Fig.~\ref{exs_pib} for an example). 
In contrast, 12.5\% of our targets are clearly 
variable and exhibit a normalized rms greater than 3 
(see Fig.~\ref{exs_pib} for an example). The remaining systems 
(13.5\%) are marginally variable with a normalized rms between 2 and 3.

In terms of planetary prospects the most interesting systems are the 
variable and marginally variable binaries. Nevertheless, the presence of a planetary
companion in orbit around the primary star is not the only way to produce 
residual-velocity variations like those observed. Alternative possibilities 
include: (i) the primary star is intrinsically variable, (ii) the system is an 
unrecognized SB2, and (iii) the system is in fact triple and the secondary is 
itself a short-period spectroscopic binary. Assuming that planets are as common
in close binaries as around single stars, we expect to find only one or two 
planets more massive than 0.5~M$_{\rm Jup}$ and with a period shorter than
$\sim$40~days in our sample. This rough estimation shows that most of the
observed residual-velocity variations are probably not related to the presence
of planetary companions, but likely stem from the binary or multiple nature of 
our targets. Therefore, to identify the few potential planet-bearing 
stars among the many variable and marginally variable systems we must find a 
way to precisely characterize the cause of the residual-velocity variations.

\subsubsection{Identifying the Origin of Residual-Velocity Variations}

\begin{figure}[htb!]
\centering
\resizebox{11.5cm}{!}{
\includegraphics{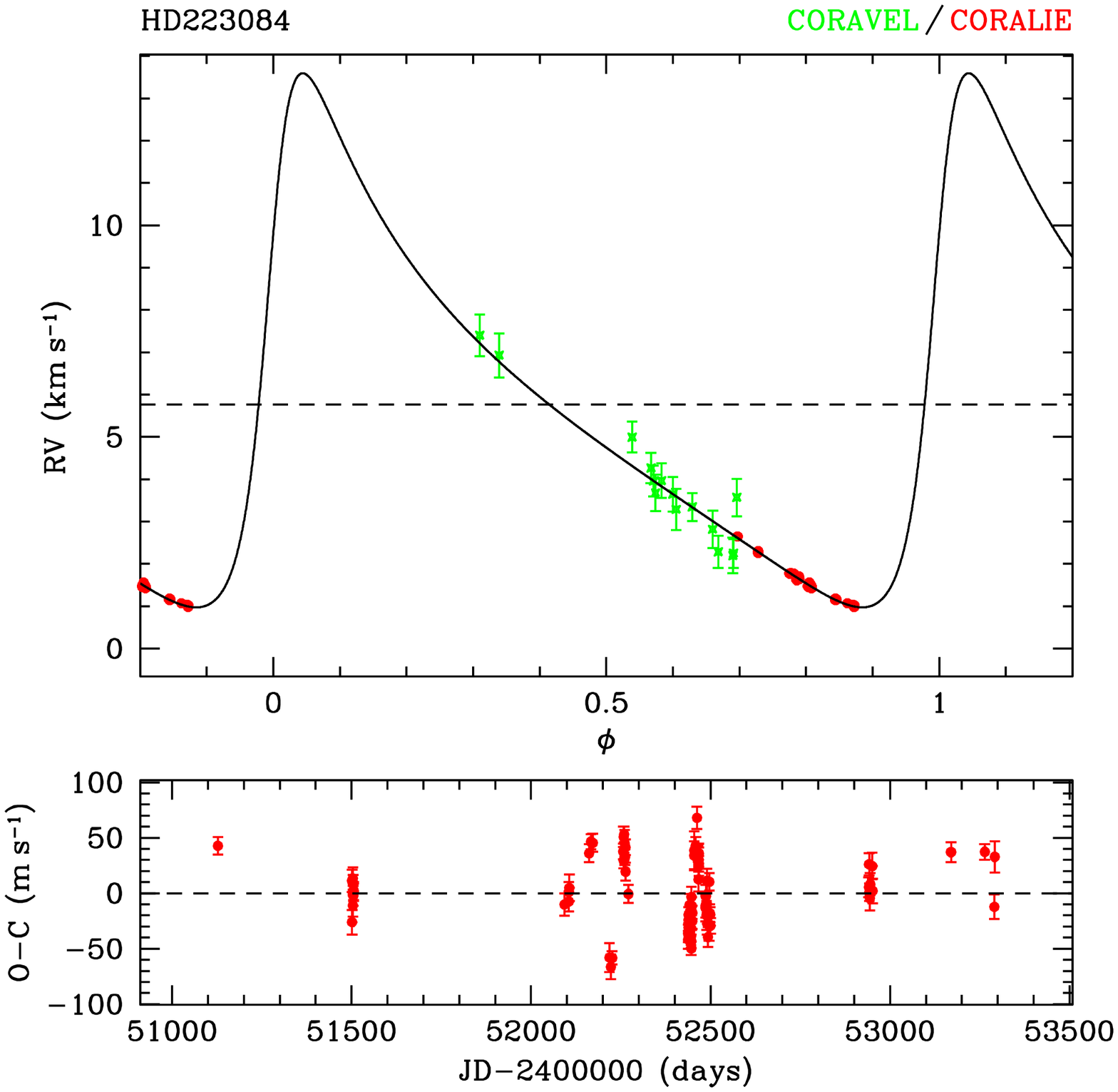}\hspace{0.1cm}
\includegraphics{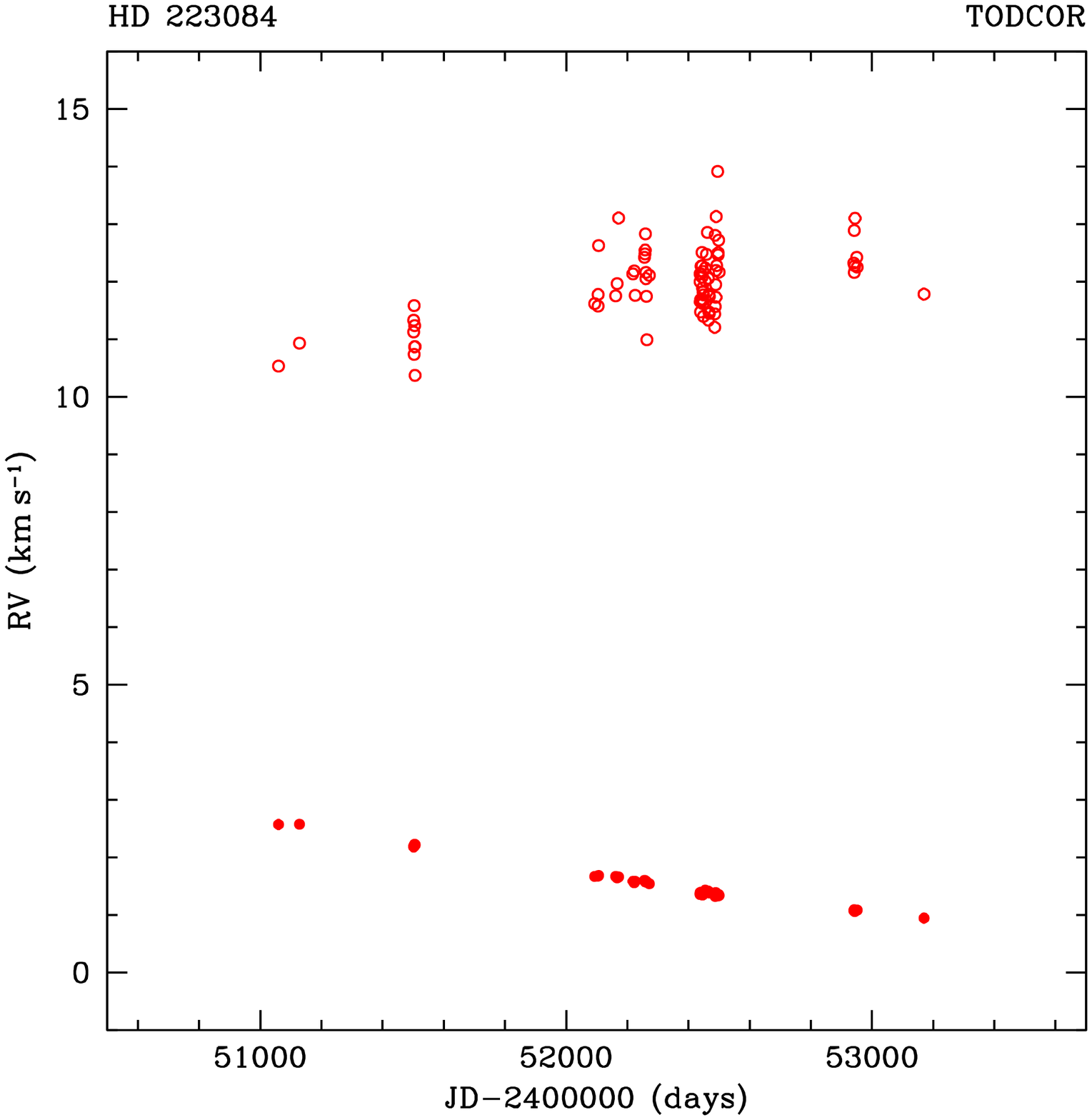}}
\resizebox{5.7cm}{!}{
\includegraphics{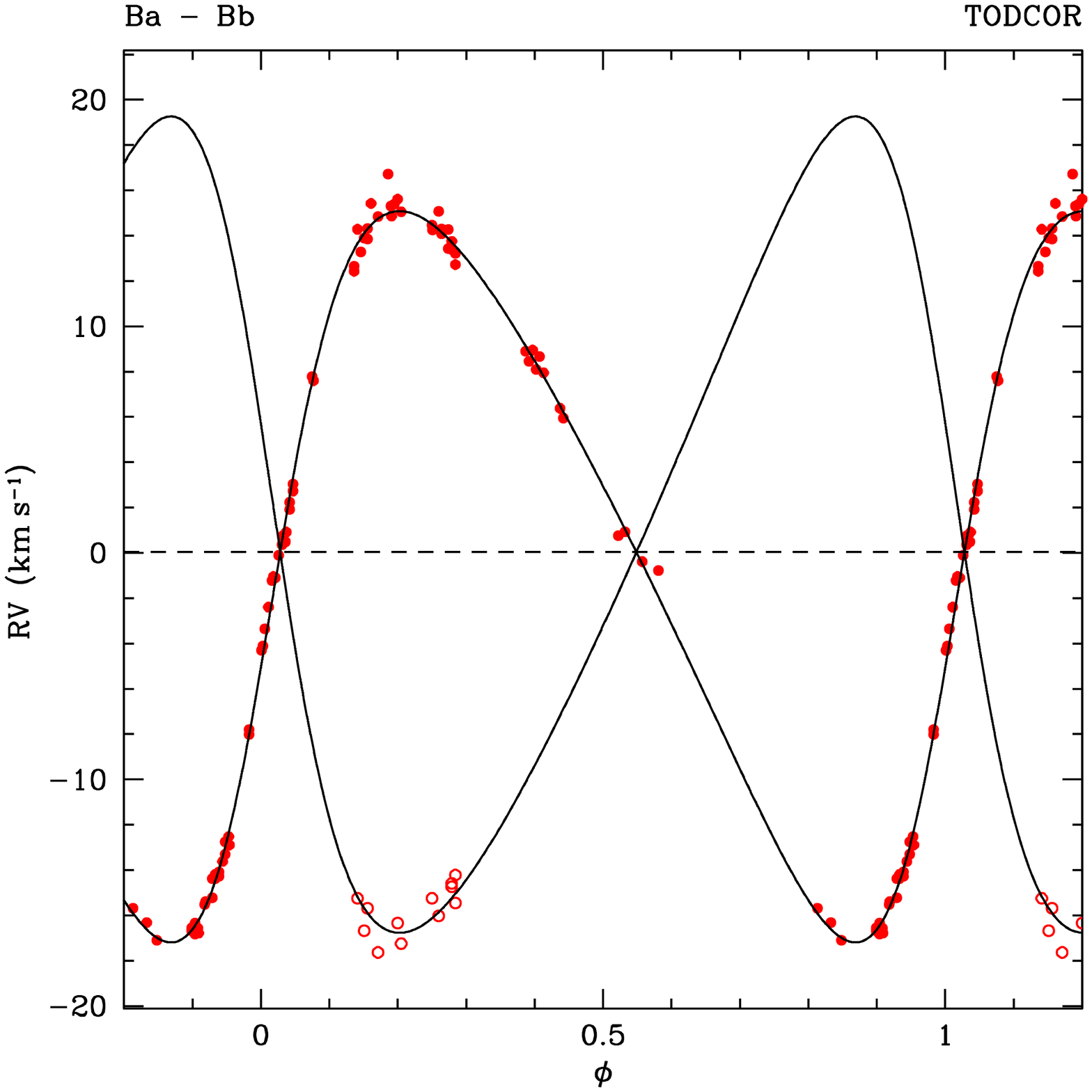}}
\caption{An example of triple system: HD\,223084. 
{\bf Top left:} CORAVEL (crosses, large error bars) and CORALIE 
(dots) velocities for HD\,223084. The binary orbit is tentative and is used 
only as a proxy to compute residual velocities. The bottom panel shows the
residual velocities (CORALIE data only). 
{\bf Top right:} TODCOR velocities for HD\,223084\,A (dots) and  
HD\,223084\,Ba (open circles) after having removed the 202-day modulation of
the Ba--Bb inner pair. 
{\bf Bottom:} SB2 orbit for  HD\,223084\,Ba (dots) and HD\,223084\,Bb
(open circles). This orbit is characterized by a period of 202~days and 
velocity semiamplitudes of 16.1~km\,s$^{-1}$ and 18~km\,s$^{-1}$ for 
components Ba and Bb, respectively.}
\label{hd223084}
\end{figure}

\begin{figure}[htb!]
\centering
\resizebox{11.5cm}{!}{
\includegraphics{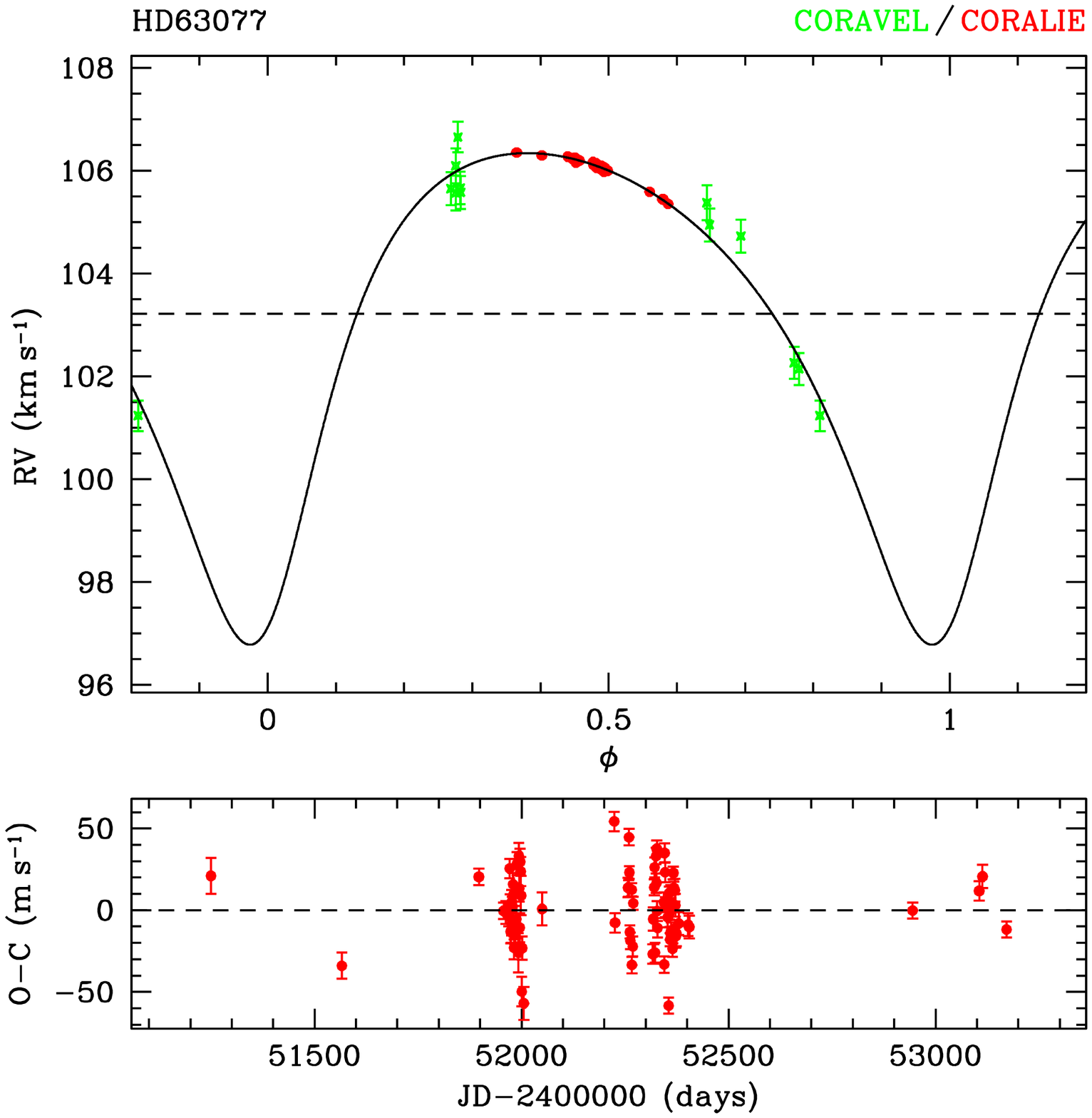}\hspace{0.1cm}
\includegraphics{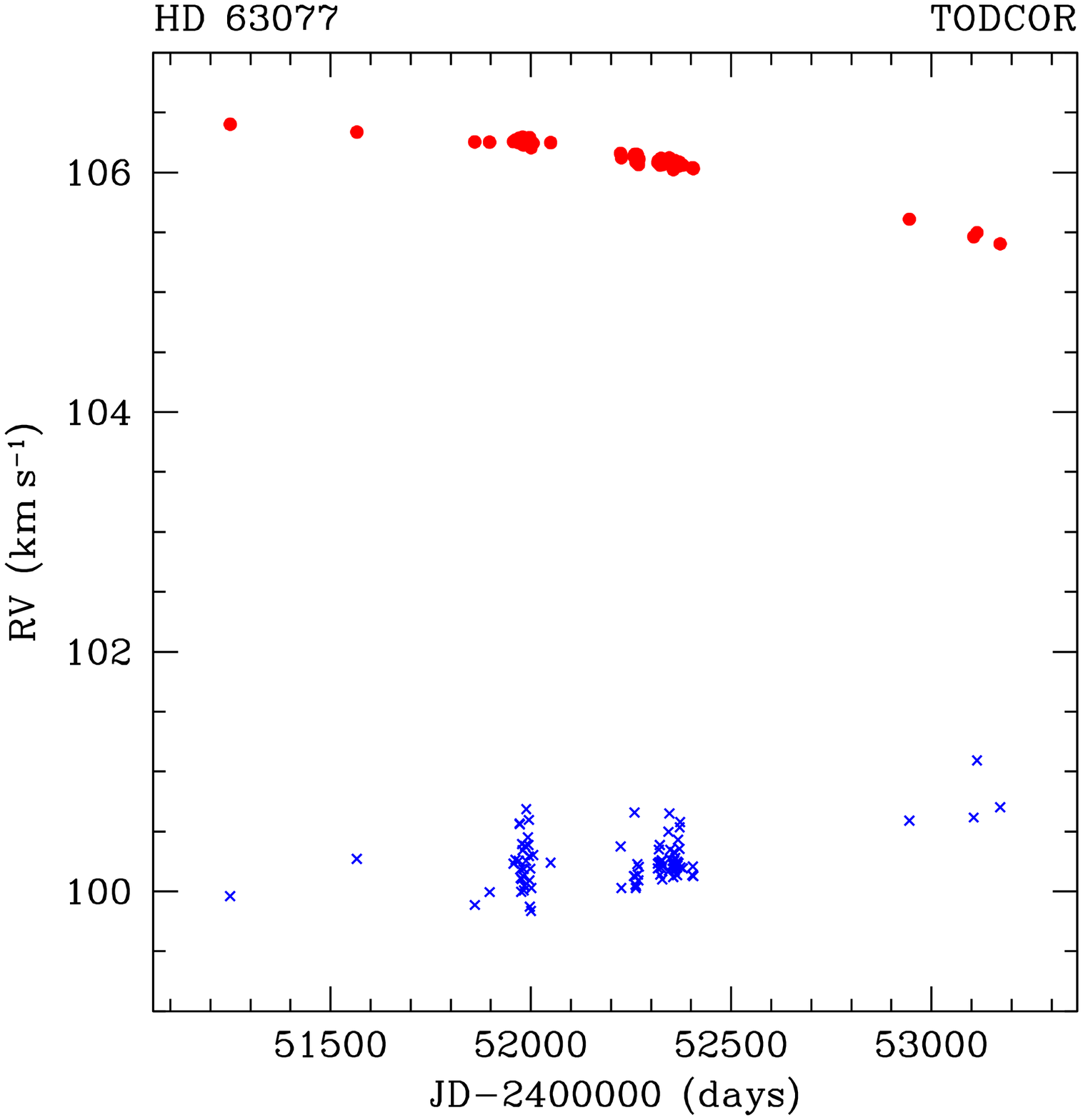}}
\caption{An example of unrecognized SB2: HD\,63077. 
{\bf Left:} CORAVEL (crosses, large error bars) and CORALIE (dots) 
velocities for HD\,63077. The binary orbit is tentative and is used only as 
a proxy to compute residual velocities. The bottom panel shows the residual
velocities (CORALIE data only). 
{\bf Right:} TODCOR velocities for HD\,63077\,A (dots) and HD\,63077\,B
(crosses). The opposite slopes of the two components are clearly seen.}
\label{hd63077}
\end{figure}

Binaries with instrinsically 
variable primaries can be identified like single active stars by considering 
the chromospheric emission flux in the Ca~\textsc{II} H and K lines. 
Using cross-correlation techniques, identifying triple systems and unrecognized 
SB2s is feasible is some instances (see e.g. 
\cite{Santos02,Eggenberger03,Eggenberger07c}), but two-dimensional 
correlation is a much more efficient tool to this purpose. 
Accordingly, we are presently 
analyzing all the variable and marginally variable systems with the 
two-dimentional algorithm TODCOR, trying to identify unrecognized SB2s and 
triple systems. 
This work is just beginning and only four variable systems have been studied 
in some detail thus far. Of these four systems, two turned out to be triple 
star systems (see Fig.~\ref{hd223084} or \cite{Eggenberger03,Eggenberger07c} 
for an example), while the two others 
turned out to be regular binaries with small relative velocities (see 
Fig.~\ref{hd63077} for an example). Not any of these four systems shows hints of the presence
of a circumprimary planet. 

\subsubsection{Preliminary General Results}

The present results from our search for circumprimary short-period planets in
SB1s show that in most of these systems (74\%) the secondary component is so
faint ($\Delta V$\,$\gtrsim$\,$6$) that it does not contribute significantly 
to the recorded flux. Doppler data of such systems can be analyzed like 
Doppler data of single stars and the precision achieved on the measurement of 
the radial velocity of the primary star is as good as for single stars. These 
SB1s can thus be included in classical radial-velocity planet searches.

In contrast, analyzing the Doppler data of the 26 SB1s that exhibit 
residual-velocity variations is not as straightforward. In many of these 
systems the secondary component (and possibly the tertiary component as well) 
significantly contributes to the recorded flux 
($\Delta V$\,$\in$\,$[\sim 3,\sim 6]$), rendering the use of two-dimensional
correlation mandatory to unambiguously identify the origin of the variations
observed and hence to search for circumprimary planets. 
Our current results do not enable us to precisely characterize our 
detection capabilities in terms of circumprimary planet searches, but we 
estimate that typical precisions on the radial velocity of the primary star 
range between 10 and 20~m\,s$^{-1}$. Although these precisions are not as good 
as for single stars, they remain good enough to search for giant planets. 

The preliminary results from our search for circumprimary giant planets in SB1s 
thus confirm that such a program has grounds for existence. So far, our survey 
has unveiled no promising planetary candidate, but the data of 22 variable
and marginally variable systems remain to be analyzed in detail with
two-dimensional correlation. Since contamination 
effects stemming from the stellar companions are likely to prevail over 
potential planetary signals, two-dimensional analyses must be completed before
concluding on the existence, or absence, of planets in our sample. All we can say at
present is that less than 22\% of the SB1s from our sample have a short-period 
($P$\,$\lesssim$\,40~days) giant ($M$\,$\gtrsim$\,0.5~M$_{\rm Jup}$) planetary 
companion. Definitive results from our survey will enable us to obtain a much 
tighter constraint.

\subsection{Searching for Planets in SB2s: the Example of HD\,188753}
\label{sb2s}

Double-lined spectroscopic binaries have not been systematically included in 
our observing programs yet, so that our experience with planet searches in 
these binaries is limited to a few systems. Among those is 
HD\,188753, a close triple star system hosting a short-period giant 
planet (a hot Jupiter) according to \cite{Konacki05}. In this section we 
present our data and analysis of 
HD\,188753, which do not confirm the existence of this hot Jupiter
\cite{Eggenberger07a}. 
Beyond the debate, HD\,188753 constitutes a concrete example of some of the 
challenges faced by Doppler searches for planets in spectroscopic binaries.

HD\,188753 has attracted much attention since July 2005 when \cite{Konacki05} 
reported the discovery of a 1.14-M$_{\rm Jup}$ planet on a 3.35-day orbit around the 
primary component of this triple star system. Aside from the 
planet, HD\,188753 consists of a primary star (HD\,188753\,A) 
orbited by a visual companion, HD\,188753\,B, which is itself a spectroscopic 
binary (i.e. HD\,188753\,B is actually made of two stellar components, 
HD\,188753\,Ba and HD\,188753\,Bb). The visual orbit of the AB pair 
is characterized by a period of 25.7~years, a semimajor axis of 12.3~AU
($0.27^{\prime\prime}$\ separation) and an eccentricity of 0.5 
\cite{Soederhjelm99}, while the spectroscopic orbit of HD\,188753\,B has a
period of 155~days \cite{Griffin77,Konacki05}. 
What renders this discovery particularly important and 
interesting is that the periastron distance of the AB pair may be small enough 
to preclude giant planet formation around HD\,188753\,A according to the 
canonical planet-formation models \cite{Nelson00,Mayer05,Boss06,Jang-Condell07}. 
The discovery of a close-in giant planet
around this star has thus been perceived as a serious challenge to
planet-formation theories, though the alternative possibility that 
HD\,188753\,A might have acquired its planet through dynamical interactions was
also pointed out \cite{Pfahl05,PortegiesZwart05}.

Following the announcement by \cite{Konacki05}, we monitored HD\,188753 during 
one year with the ELODIE spectrograph, gathering a total of 48 
spectra. The cross-correlation at the telescope immediately revealed the 
double-lined nature of HD\,188753, the cross-correlation function consisting of 
two blended 
features corresponding to components A and Ba, respectively. The contribution 
of the third component (Bb) to the total flux is quite 
modest, so that the system can basically be considered as a double-lined 
spectroscopic binary. 

\begin{figure}[htb!]
\centering
\resizebox{11.5cm}{!}{
\includegraphics{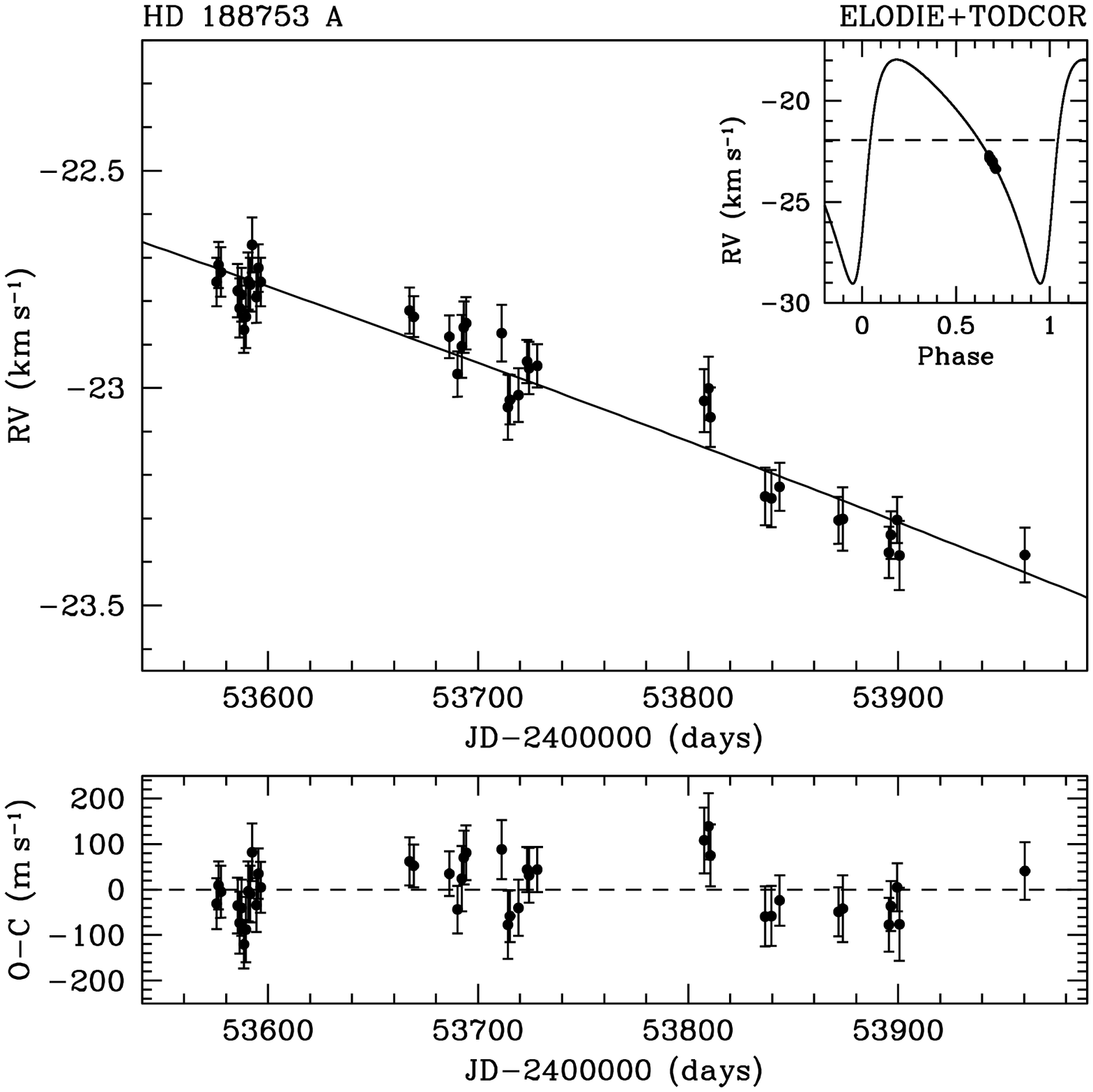}\hspace{0.1cm}
\includegraphics{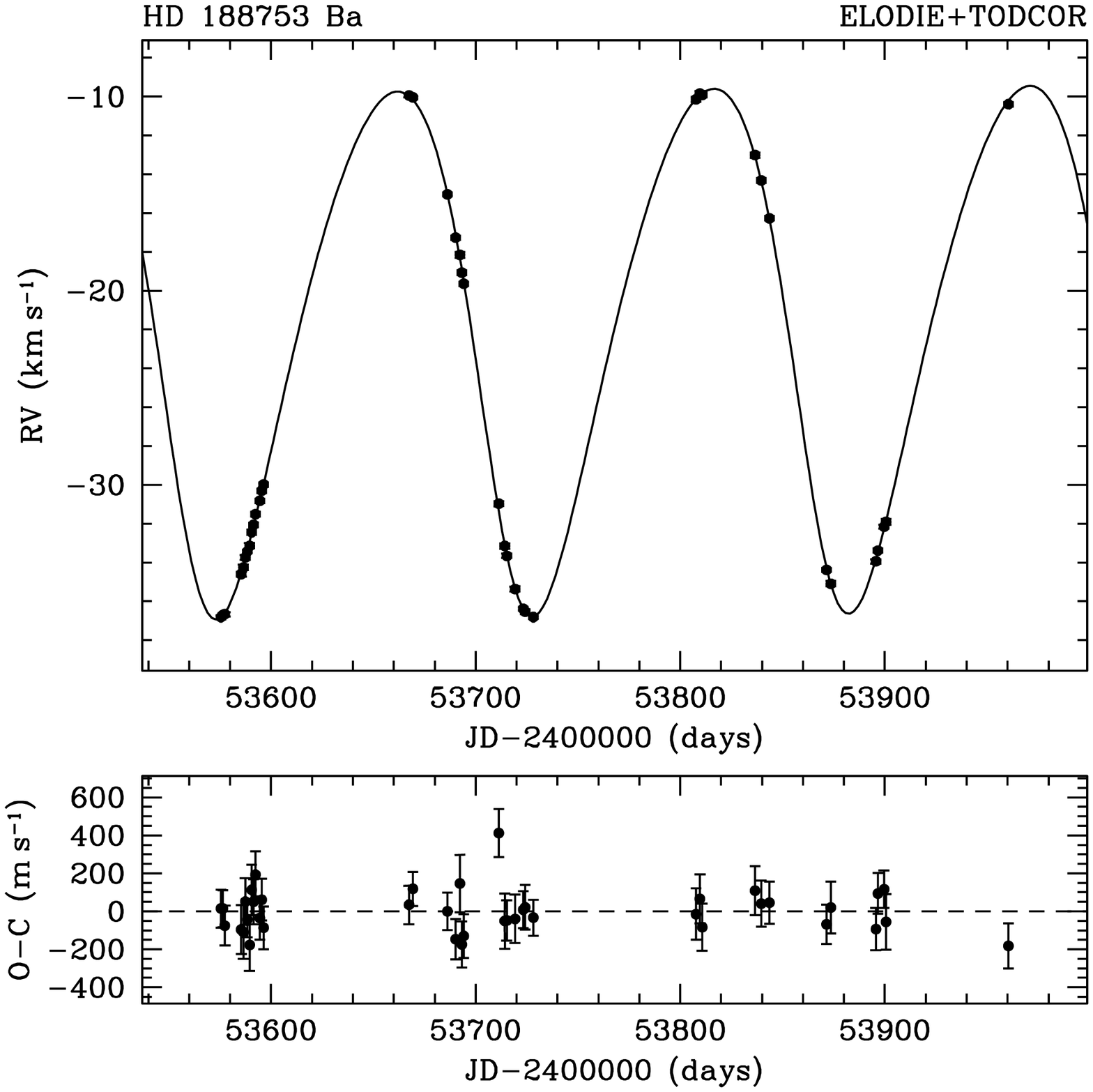}}
\resizebox{11.5cm}{!}{
\includegraphics{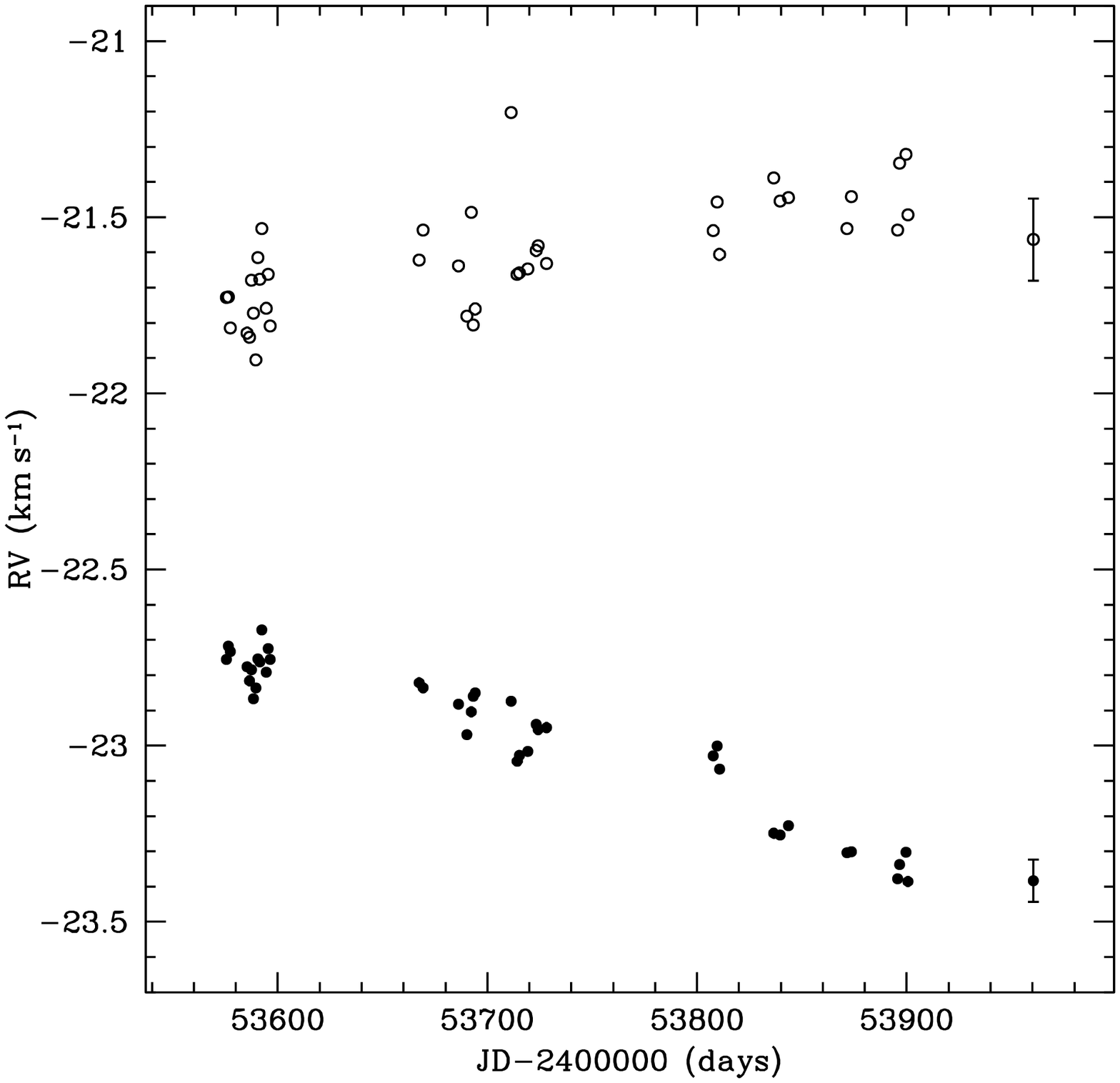}\hspace{0.1cm}
\includegraphics{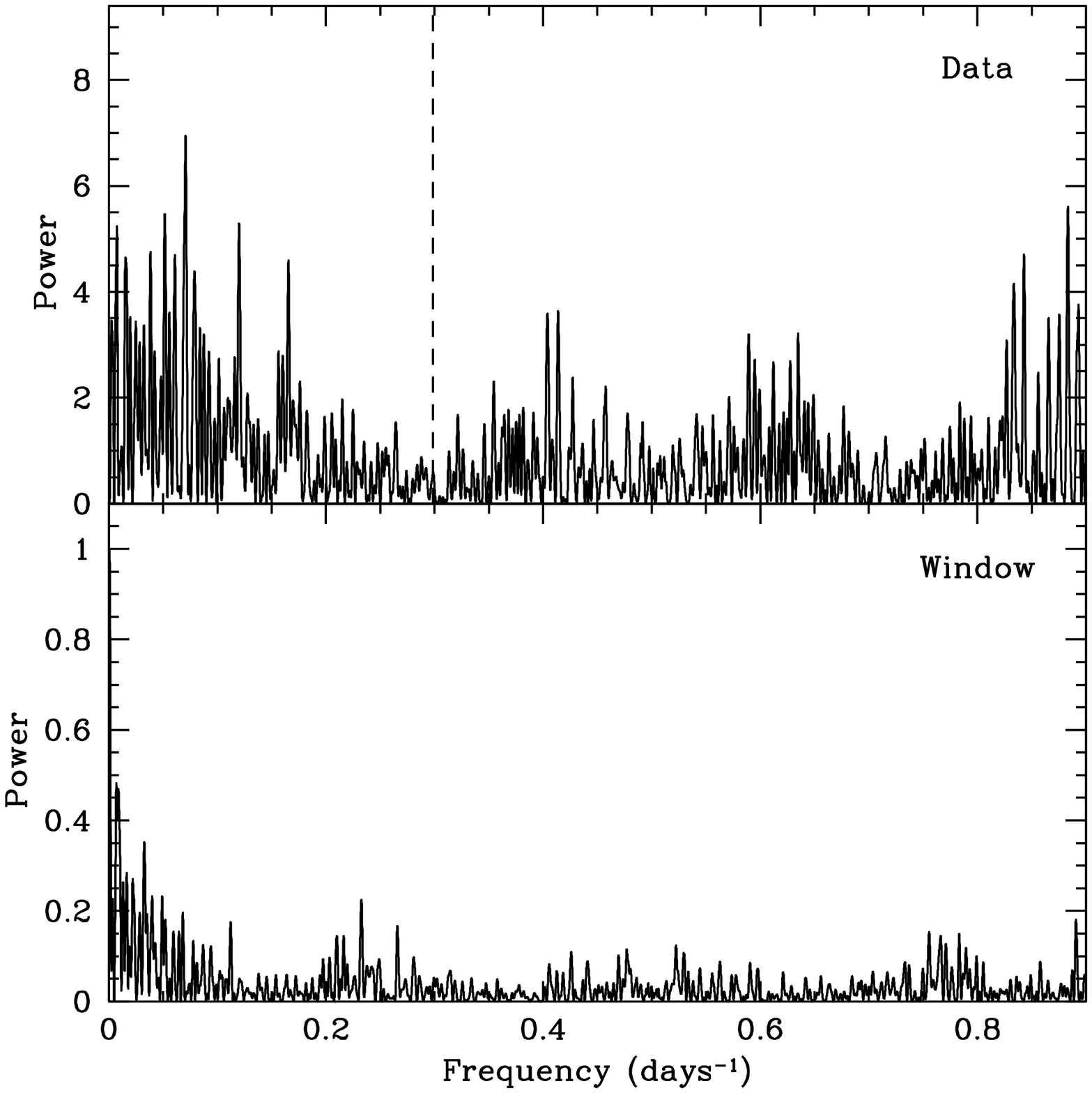}}
\caption{{\bf Top:} Radial velocities and orbital solutions for 
HD\,188753\,A (left) and HD\,188753\,Ba (right). For component A, the solid 
line represents the 25.7-year orbital motion of the visual pair shown in
full in the inset. For component Ba the orbital solution corresponds to the
155-day modulation and it includes a linear drift to take the 25.7-year orbital 
motion into account. 
{\bf Bottom left:} Radial velocities for HD\,188753\,A (dots) and 
for HD\,188753\,Ba after having removed the 155-day modulation (open circles). 
For the sake of clarity, only a typical error bar is displayed on the last
measurement of each component. 
{\bf Bottom right:} Lomb-Scargle periodogram of the residuals around the 
25.7-year orbital motion for HD\,188753\,A. The 1\% false alarm probability 
corresponds to a power of 9.4 and is represented by the top of the box. The 
dashed line denotes the frequency of the planetary signal reported by 
\cite{Konacki05}.}
\label{hd188753}
\end{figure}

Given the double-lined nature of HD\,188753 and the strong line blending, we 
derived the radial velocities of HD\,188753\,A and HD\,188753\,Ba 
using the TODCOR algorithm. These velocities are displayed in 
Fig.~\ref{hd188753}. Our results for HD\,188753\,Ba confirm that it is indeed
a spectroscopic binary with a period of 155~days. 
As to HD\,188753\,A, the dominant motion seen in our data is a steady 
decrease in velocity, fully consistent with the 25.7-year orbital motion 
of the AB pair. However, our radial velocities show no sign of the additional 
3.35-day planetary signal reported by 
\cite{Konacki05}. Instead, the residuals around the long-period drift are 
basically noise (Fig.~\ref{hd188753}) and the rms of 60~m\,s$^{-1}$ can be 
interpreted as the precision we achieve on the measurement of the radial 
velocity of this star. Monte Carlo simulations run to check our
ability to detect the potential planet around
HD\,188753\,A show that we had both the precision and the temporal sampling 
required to detect a planetary signal like the one reported by 
\cite{Konacki05}. On that basis, we conclude that our data show no evidence of 
a 1.14\,M$_{\rm Jup}$ on a 3.35-day orbit around HD\,188753\,A.

While three planets have been discovered so far in binaries with a 
separation of $\sim$$20$~AU (Table~\ref{pib_table}), the planet found around 
HD\,188753\,A was the only planet known to reside in a tighter system.  
Therefore, the removal of HD\,188753\,Ab from the list of planetary candidates 
not only eliminates a potential source of difficulty for theorists, but also 
brings further support to the idea that the ``limit'' at 20~AU might
be associated with a minimum separation for considering that a binary
possibly harbors a giant planet. 

Our experience with SB2 systems is as yet too limited to enable us to 
characterize our detection capabilities in terms of planet searches around the
primary stars in these binaries. Nonetheless, in view of the results presented
in Sect.~\ref{sb1s}, the precision of 60~m\,s$^{-1}$ obtained on the radial
velocity of HD\,188753\,A looks abnormally poor. Further investigations are 
underway to specify the main factor that limits our current precision on the 
radial velocity of HD\,188753\,A. In any way, the triple nature of HD\,188753 
allows us to hope that better precisions may be achieved for the primary
components of true SB2s.


\section{Conclusion and Perspectives}
\label{conclusion}

Over the past five years, binaries have become increasingly interesting targets
in terms of planet searches. One the one hand, Doppler surveys have shown that 
giant planets exist not only in wide binaries but also in the much closer
spectroscopic binaries, raising the possibility that planets 
might be common in binaries and multiple stars. On the other hand, theoretical
studies have shown that the presence of a close ($\lesssim$100-120~AU) stellar
companion affects the formation and subsequent evolution of circumstellar giant 
planets, leaving some imprints in the occurrence, characteristics, and 
properties of the planets 
residing in these systems. The study of circumstellar planets in close binaries 
might thus provide a unique means to probe the main formation mechanism for 
giant planets and to bring observational constraints for planet-formation 
models. 

Imaging surveys searching for stellar companions to planet-bearing stars
have been very successful, yielding a precise  
characterization of the multiplicity status of more than half of the known 
planet-host stars. More importantly, our NACO survey has provided us not only 
with the multiplicity status of $\sim$70 planet-host 
stars, but also with the multiplicity status of the same number of comparison 
stars showing the least possible evidence for planetary companions and 
affected by the same selection effects than planet-host stars. A preliminary 
statistical analysis of this unprecedented data set brings the first 
observational evidence that the occurrence of planets is reduced in binaries 
closer than $\sim$120~AU. Given our present knowledge of planet-formation 
mechanisms, at
least two different explanations can be put forward to explain this result:
either disk instability is a viable formation mechanism that accounts for the
existence of a significant number of the planets known presently, or core
accretion is the main formation channel but its efficiency is reduced in
binaries closer than $\sim$120~AU. Differentiating between these two
possibilities will require some additional work, both on the theoretical and on 
the observational sides. Yet, the important point to notice is that 
observations have  
caught up with theoretical studies on the investigation of the impact of 
stellar duplicity on giant planet formation, meaning that some theoretical 
predictions can now be confronted with observational results.

The recent discoveries from imaging surveys have somewhat decreased the 
statistical significance of the emerging trends suggesting that short-period
planets found in binaries and multiple stars possess distinctive characteristics
and properties compared to their counterparts orbiting single stars. The most
robust feature in this respect is still the observation that the few most 
massive short-period planets all orbit the components of binaries or triple 
stars. Nonetheless, such planets are still sparse (no new discovery since 
2003) and even the most recent statistical studies remain affected by the uncertain 
multiplicity status of a large number of planet-host stars. The combined 
results from our NACO and PUEO surveys will remove this last uncertainty to a 
large extent, allowing for a major reinvestigation of possible differences in 
the eccentricity distributions of planet-host stars found in
binaries and around single stars. 

During the last few years, significant effort has been put into extending 
radial-velocity planet searches to spectroscopic and close visual binaries. 
Doppler surveys dedicated to these close systems have 
proven that the main difficulties associated with the recording of composite 
spectra can be overcome by the use of two-dimensional correlation algorithms. 
In a general way, planet searches in close binaries are still in their early 
phases and only partial results are available yet. Nonetheless, current results 
demonstrate that Doppler searches for giant planets in single-lined and in some 
types of double-lined spectroscopic binaries are technically feasible. 
Searches for lower mass planets in some types of close binaries are not 
excluded a priori, but developing the technique further will first require a 
better understanding of the main limiting factors. 

Final results from the presently ongoing planet searches in spectroscopic 
binaries are awaited with
great interest for several reasons. First, these surveys constitute the only 
current possibility to probe the occurrence of giant planets in the closest 
binaries and to characterize the closest systems susceptible of hosting 
circumstellar giant planets. In this context, Doppler searches for planets in 
spectroscopic 
binaries will provide us with stronger constraints on the reality of the 
20-AU ``limit'' and on its possible interpretation as a minimum separation for 
considering that a binary possibly harbors a giant planet. Second, the outcomes 
from planet searches in spectroscopic binaries will allow us to quantify the 
occurrence of giant planets in binaries closer than $\sim$35~AU. 
This, in turn, will nicely complement the results from our NACO and PUEO 
surveys. Gathering together the observational results from our imaging and
radial-velocity programs might then provide us with 
some constraints as to whether most giant planets found in binaries closer than 
$\sim$50~AU actually formed in these systems, or were deposited at their present location 
through dynamical interactions. Finally, radial-velocity planets searches are
the best tool to expand the size of the still limited sample of planets 
residing in binaries and multiple stars. 

As planet searches progress, the conviction that planets are common objects in 
the universe continually strengthen. The discovery of giant planets in 
environments previously considered as relatively hostile to their existence 
(spectroscopic binaries, pulsars, ...) has contributed to this development, 
showing that planet formation is not as easily inhibited as originally thought. 
In addition to the encouraging results obtained thus far, the expectation that 
terrestrial planets form alongside their Jovian counterparts suggests that 
discoveries are limited by instrumental sensitivity rather than the 
availability of planets. Even if the presence of a close stellar companion 
lowers the efficiency of planet formation, theoretical studies support the 
existence of circumstellar terrestrial planets in many types of binaries. On the
other hand, circumbinary planets are also expected to exist and searches for
circumbinary planets offer a still unexplored field of investigation for planet
hunters. Planet searches in and around binaries are thus not only meaningful, 
but also desirable in view of the potential information they can yield on the 
overall frequency of planets and on the processes underlying planet formation 
and evolution.

\begin{acknowledgement}
We thank A.-M.~Lagrange for helpful comments on our manuscript. 
\end{acknowledgement}

%
%
%
 \bibliographystyle{plain}
 \bibliography{eggenberger}

\begin{thebibliography}{10}

\bibitem{Artymowicz94}
P.~{Artymowicz} and S.~H. {Lubow}.
\newblock {Dynamics of binary-disk interaction. 1: Resonances and disk gap
  sizes}.
\newblock {\em ApJ}, 421:651--667, February 1994.

\bibitem{Baranne79}
A.~{Baranne}, M.~{Mayor}, and J.~L. {Poncet}.
\newblock {CORAVEL - A new tool for radial velocity measurements}.
\newblock {\em Vistas in Astronomy}, 23:279--316, 1979.

\bibitem{Baranne96}
A.~{Baranne}, D.~{Queloz}, M.~{Mayor}, G.~{Adrianzyk}, G.~{Knispel},
  D.~{Kohler}, D.~{Lacroix}, J.-P. {Meunier}, G.~{Rimbaud}, and A.~{Vin}.
\newblock {ELODIE: A spectrograph for accurate radial velocity measurements.}
\newblock {\em A\&AS}, 119:373--390, October 1996.

\bibitem{Boss06}
A.~P. {Boss}.
\newblock {Gas Giant Protoplanets Formed by Disk Instability in Binary Star
  Systems}.
\newblock {\em ApJ}, 641:1148--1161, April 2006.

\bibitem{Butler97}
R.~P. {Butler}, G.~W. {Marcy}, E.~{Williams}, H.~{Hauser}, and P.~{Shirts}.
\newblock {Three New ''51 Pegasi--Type'' Planets}.
\newblock {\em ApJL}, 474:L115, January 1997.

\bibitem{Chauvin06}
G.~{Chauvin}, A.-M. {Lagrange}, S.~{Udry}, T.~{Fusco}, F.~{Galland}, D.~{Naef},
  J.-L. {Beuzit}, and M.~{Mayor}.
\newblock {Probing long-period companions to planetary hosts. VLT and CFHT near
  infrared coronographic imaging surveys}.
\newblock {\em A\&A}, 456:1165--1172, September 2006.

\bibitem{Cochran97}
W.~D. {Cochran}, A.~P. {Hatzes}, R.~P. {Butler}, and G.~W. {Marcy}.
\newblock {The Discovery of a Planetary Companion to 16 Cygni B}.
\newblock {\em ApJ}, 483:457, July 1997.

\bibitem{Delfosse04}
X.~{Delfosse}, J.-L. {Beuzit}, L.~{Marchal}, X.~{Bonfils}, C.~{Perrier},
  D.~{S{\'e}gransan}, S.~{Udry}, M.~{Mayor}, and T.~{Forveille}.
\newblock {M dwarfs binaries: Results from accurate radial velocities and high
  angular resolution observations}.
\newblock In R.~W. {Hilditch}, H.~{Hensberge}, and K.~{Pavlovski}, editors,
  {\em ASP Conf. Ser. 318: Spectroscopically and Spatially Resolving the
  Components of the Close Binary Stars}, pages 166--174, December 2004.

\bibitem{Desidera07}
S.~{Desidera} and M.~{Barbieri}.
\newblock {Properties of planets in binary systems. The role of binary
  separation}.
\newblock {\em A\&A}, 462:345--353, January 2007.

\bibitem{Duquennoy91}
A.~{Duquennoy} and M.~{Mayor}.
\newblock {Multiplicity among solar-type stars in the solar neighbourhood. II -
  Distribution of the orbital elements in an unbiased sample}.
\newblock {\em A\&A}, 248:485--524, August 1991.

\bibitem{Duquennoy91b}
A.~{Duquennoy}, M.~{Mayor}, and J.-L. {Halbwachs}.
\newblock {Multiplicity among solar type stars in the solar neighbourhood. I -
  CORAVEL radial velocity observations of 291 stars}.
\newblock {\em A\&AS}, 88:281--324, May 1991.

\bibitem{Durisen07}
R.~H. {Durisen}, A.~P. {Boss}, L.~{Mayer}, A.~F. {Nelson}, T.~{Quinn}, and
  W.~K.~M. {Rice}.
\newblock {Gravitational Instabilities in Gaseous Protoplanetary Disks and
  Implications for Giant Planet Formation}.
\newblock In B.~{Reipurth}, D.~{Jewitt}, and K.~{Keil}, editors, {\em
  Protostars and Planets V}, pages 607--622, 2007.

\bibitem{Eggenberger04b}
A.~{Eggenberger}, J.-L. {Halbwachs}, S.~{Udry}, and M.~{Mayor}.
\newblock {Statistical properties of an unbiased sample of F7-K binaries:
  towards the long-period systems}.
\newblock In C.~{Allen} and C.~{Scarfe}, editors, {\em Revista Mexicana de
  Astronomia y Astrofisica Conference Series}, pages 28--32, August 2004.

\bibitem{Eggenberger07b}
A.~{Eggenberger}, S.~{Udry}, G.~{Chauvin}, J.-L. {Beuzit}, A.-M. {Lagrange},
  D.~{S\'egransan}, and M.~{Mayor}.
\newblock {The impact of stellar duplicity on planet formation and evolution I.
  The multiplicity status of nearby stars with and without planets probed with
  VLT/NACO}.
\newblock {\em A\&A}, 2007.
\newblock submitted.

\bibitem{Eggenberger03}
A.~{Eggenberger}, S.~{Udry}, and M.~{Mayor}.
\newblock {Planets in Binaries}.
\newblock In {\em ASP Conf. Ser. 294: Scientific Frontiers in Research on
  Extrasolar Planets}, pages 43--46, 2003.

\bibitem{Eggenberger04}
A.~{Eggenberger}, S.~{Udry}, and M.~{Mayor}.
\newblock {Statistical properties of exoplanets. III. Planet properties and
  stellar multiplicity}.
\newblock {\em A\&A}, 417:353--360, April 2004.

\bibitem{Eggenberger04c}
A.~{Eggenberger}, S.~{Udry}, M.~{Mayor}, J.-L. {Beuzit}, A.~M. {Lagrange}, and
  G.~{Chauvin}.
\newblock {Detection and Properties of Extrasolar Planets in Double and
  Multiple Star Systems}.
\newblock In J.~{Beaulieu}, A.~{Lecavelier Des Etangs}, and C.~{Terquem},
  editors, {\em ASP Conf. Ser. 321: Extrasolar Planets: Today and Tomorrow},
  page~93, December 2004.

\bibitem{Eggenberger07c}
A.~{Eggenberger}, S.~{Udry}, M.~{Mayor}, G.~{Chauvin}, B.~{Markus}, J.-L.
  {Beuzit}, A.~M. {Lagrange}, T.~{Mazeh}, S.~{Zucker}, and D.~{S\'egransan}.
\newblock {Extrasolar Planets in Double and Multiple Stellar Systems}.
\newblock In {\em Multiple Stars Across the H-R Diagram, ESO Astrophysic
  Symposia, Edited by S. Hubrig, M. Petr-Gotzens and A. Tokovinin}, 2007.
\newblock in press, available at
  www.eso.org/gen-fac/meetings/ms2005/eggenberger.pdf.

\bibitem{Eggenberger07a}
A.~{Eggenberger}, S.~{Udry}, T.~{Mazeh}, Y.~{Segal}, and M.~{Mayor}.
\newblock {No evidence of a hot Jupiter around HD 188753 A}.
\newblock {\em A\&A}, 2007.
\newblock in press.

\bibitem{Fatuzzo06}
M.~{Fatuzzo}, F.~C. {Adams}, R.~{Gauvin}, and E.~M. {Proszkow}.
\newblock {A Statistical Stability Analysis of Earth-like Planetary Orbits in
  Binary Systems}.
\newblock {\em PASP}, 118:1510--1527, November 2006.

\bibitem{Fischer92}
D.~A. {Fischer} and G.~W. {Marcy}.
\newblock {Multiplicity among M dwarfs}.
\newblock {\em ApJ}, 396:178--194, September 1992.

\bibitem{Griffin77}
R.~F. {Griffin}.
\newblock {The multiple star HD 188753 (ADS 13125)}.
\newblock {\em The Observatory}, 97:15--18, February 1977.

\bibitem{Halbwachs05}
J.~L. {Halbwachs}, M.~{Mayor}, and S.~{Udry}.
\newblock {Statistical properties of exoplanets. IV. The period-eccentricity
  relations of exoplanets and of binary stars}.
\newblock {\em A\&A}, 431:1129--1137, March 2005.

\bibitem{Halbwachs03}
J.~L. {Halbwachs}, M.~{Mayor}, S.~{Udry}, and F.~{Arenou}.
\newblock {Multiplicity among solar-type stars. III. Statistical properties of
  the F7-K binaries with periods up to 10 years}.
\newblock {\em A\&A}, 397:159--175, January 2003.

\bibitem{Holman97}
M.~{Holman}, J.~{Touma}, and S.~{Tremaine}.
\newblock {Chaotic variations in the eccentricity of the planet orbiting 16
  Cygni B}.
\newblock {\em Nature}, 386:254--256, March 1997.

\bibitem{Holman99}
M.~J. {Holman} and P.~A. {Wiegert}.
\newblock {Long-Term Stability of Planets in Binary Systems}.
\newblock {\em AJ}, 117:621--628, January 1999.

\bibitem{Innanen97}
K.~A. {Innanen}, J.~Q. {Zheng}, S.~{Mikkola}, and M.~J. {Valtonen}.
\newblock {The Kozai Mechanism and the Stability of Planetary Orbits in Binary
  Star Systems}.
\newblock {\em AJ}, 113:1915, May 1997.

\bibitem{Jang-Condell07}
H.~{Jang-Condell}.
\newblock {Constraints on the Formation of the Planet in HD 188753}.
\newblock {\em ApJ}, 654:641--649, January 2007.

\bibitem{Jones06}
H.~R.~A. {Jones}, R.~P. {Butler}, C.~G. {Tinney}, G.~W. {Marcy}, B.~D.
  {Carter}, A.~J. {Penny}, C.~{McCarthy}, and J.~{Bailey}.
\newblock {High-eccentricity planets from the Anglo-Australian Planet Search}.
\newblock {\em MNRAS}, 369:249--256, June 2006.

\bibitem{Kley00}
W.~{Kley}.
\newblock {Evolution of an embedded Planet in a Binary System}.
\newblock In {\em IAU Symposium}, page 211P, 2000.

\bibitem{Konacki05}
M.~{Konacki}.
\newblock {An extrasolar giant planet in a close triple-star system}.
\newblock {\em Nature}, 436:230--233, July 2005.

\bibitem{Konacki05b}
M.~{Konacki}.
\newblock {Precision Radial Velocities of Double-lined Spectroscopic Binaries
  with an Iodine Absorption Cell}.
\newblock {\em ApJ}, 626:431--438, June 2005.

\bibitem{Lissauer07}
J.~J. {Lissauer} and D.~J. {Stevenson}.
\newblock {Formation of Giant Planets}.
\newblock In B.~{Reipurth}, D.~{Jewitt}, and K.~{Keil}, editors, {\em
  Protostars and Planets V}, pages 591--606, 2007.

\bibitem{Marcy05b}
G.~{Marcy}, R.~P. {Butler}, D.~{Fischer}, S.~{Vogt}, J.~T. {Wright}, C.~G.
  {Tinney}, and H.~R.~A. {Jones}.
\newblock {Observed Properties of Exoplanets: Masses, Orbits, and
  Metallicities}.
\newblock {\em Progress of Theoretical Physics Supplement}, 158:24--42, 2005.

\bibitem{Marcy05}
G.~W. {Marcy}, R.~P. {Butler}, S.~S. {Vogt}, D.~A. {Fischer}, G.~W. {Henry},
  G.~{Laughlin}, J.~T. {Wright}, and J.~A. {Johnson}.
\newblock {Five New Extrasolar Planets}.
\newblock {\em ApJ}, 619:570--584, January 2005.

\bibitem{Mayer05}
L.~{Mayer}, J.~{Wadsley}, T.~{Quinn}, and J.~{Stadel}.
\newblock {Gravitational instability in binary protoplanetary discs: new
  constraints on giant planet formation}.
\newblock {\em MNRAS}, 363:641--648, October 2005.

\bibitem{Mazeh97}
T.~{Mazeh}, Y.~{Krymolowski}, and G.~{Rosenfeld}.
\newblock {The High Eccentricity of the Planet Orbiting 16 Cygni B}.
\newblock {\em ApJl}, 477:L103, March 1997.

\bibitem{Mugrauer06}
M.~{Mugrauer}, R.~{Neuh{\"a}user}, T.~{Mazeh}, E.~{Guenther},
  M.~{Fern{\'a}ndez}, and C.~{Broeg}.
\newblock {A search for wide visual companions of exoplanet host stars: The
  Calar Alto Survey}.
\newblock {\em Astronomische Nachrichten}, 327:321, May 2006.

\bibitem{Mugrauer05a}
M.~{Mugrauer}, R.~{Neuh{\"a}user}, A.~{Seifahrt}, T.~{Mazeh}, and
  E.~{Guenther}.
\newblock {Four new wide binaries among exoplanet host stars}.
\newblock {\em A\&A}, 440:1051--1060, September 2005.

\bibitem{Nagasawa07}
M.~{Nagasawa}, E.~W. {Thommes}, S.~J. {Kenyon}, B.~C. {Bromley}, and D.~N.~C.
  {Lin}.
\newblock {The Diverse Origins of Terrestrial-Planet Systems}.
\newblock In B.~{Reipurth}, D.~{Jewitt}, and K.~{Keil}, editors, {\em
  Protostars and Planets V}, pages 639--654, 2007.

\bibitem{Nelson00}
A.~F. {Nelson}.
\newblock {Planet Formation is Unlikely in Equal-Mass Binary Systems with A
  \~{} 50 AU}.
\newblock {\em ApJL}, 537:L65--L68, July 2000.

\bibitem{Patience02}
J.~{Patience}, R.~J. {White}, A.~M. {Ghez}, C.~{McCabe}, I.~S. {McLean}, J.~E.
  {Larkin}, L.~{Prato}, S.~S. {Kim}, J.~P. {Lloyd}, M.~C. {Liu}, J.~R.
  {Graham}, B.~A. {Macintosh}, D.~T. {Gavel}, C.~E. {Max}, B.~J. {Bauman},
  S.~S. {Olivier}, P.~{Wizinowich}, and D.~S. {Acton}.
\newblock {Stellar Companions to Stars with Planets}.
\newblock {\em ApJ}, 581:654--665, December 2002.

\bibitem{Pepe02}
F.~{Pepe}, M.~{Mayor}, F.~{Galland}, D.~{Naef}, D.~{Queloz}, N.~C. {Santos},
  S.~{Udry}, and M.~{Burnet}.
\newblock {The CORALIE survey for southern extra-solar planets VII. Two
  short-period Saturnian companions to HD 108147 and HD 168746}.
\newblock {\em A\&A}, 388:632--638, June 2002.

\bibitem{Pepe04}
F.~{Pepe}, M.~{Mayor}, D.~{Queloz}, W.~{Benz}, X.~{Bonfils}, F.~{Bouchy}, G.~L.
  {Curto}, C.~{Lovis}, D.~{M{\'e}gevand}, C.~{Moutou}, D.~{Naef},
  G.~{Rupprecht}, N.~C. {Santos}, J.-P. {Sivan}, D.~{Sosnowska}, and S.~{Udry}.
\newblock {The HARPS search for southern extra-solar planets. I. HD 330075 b: A
  new ``hot Jupiter''}.
\newblock {\em A\&A}, 423:385--389, August 2004.

\bibitem{Perrier03}
C.~{Perrier}, J.-P. {Sivan}, D.~{Naef}, J.~L. {Beuzit}, M.~{Mayor},
  D.~{Queloz}, and S.~{Udry}.
\newblock {The ELODIE survey for northern extra-solar planets. I. Six new
  extra-solar planet candidates}.
\newblock {\em A\&A}, 410:1039--1049, November 2003.

\bibitem{Pfahl05}
E.~{Pfahl}.
\newblock {Cluster Origin of the Triple Star HD 188753 and Its Planet}.
\newblock {\em ApJL}, 635:L89--L92, December 2005.

\bibitem{Pfahl06}
E.~{Pfahl} and M.~{Muterspaugh}.
\newblock {Impact of Stellar Dynamics on the Frequency of Giant Planets in
  Close Binaries}.
\newblock {\em ApJ}, 652:1694--1697, December 2006.

\bibitem{Pichardo05}
B.~{Pichardo}, L.~S. {Sparke}, and L.~A. {Aguilar}.
\newblock {Circumstellar and circumbinary discs in eccentric stellar binaries}.
\newblock {\em MNRAS}, 359:521--530, May 2005.

\bibitem{PortegiesZwart05}
S.~F. {Portegies Zwart} and S.~L.~W. {McMillan}.
\newblock {Planets in Triple Star Systems: The Case of HD 188753}.
\newblock {\em ApJL}, 633:L141--L144, November 2005.

\bibitem{Queloz00}
D.~{Queloz}, M.~{Mayor}, L.~{Weber}, A.~{Bl{\'e}cha}, M.~{Burnet},
  B.~{Confino}, D.~{Naef}, F.~{Pepe}, N.~{Santos}, and S.~{Udry}.
\newblock {The CORALIE survey for southern extra-solar planets. I. A planet
  orbiting the star Gliese 86}.
\newblock {\em A\&A}, 354:99--102, February 2000.

\bibitem{Raghavan06}
D.~{Raghavan}, T.~J. {Henry}, B.~D. {Mason}, J.~P. {Subasavage}, W.-C. {Jao},
  T.~D. {Beaulieu}, and N.~C. {Hambly}.
\newblock {Two Suns in The Sky: Stellar Multiplicity in Exoplanet Systems}.
\newblock {\em ApJ}, 646:523--542, July 2006.

\bibitem{Santos03}
N.~C. {Santos}, G.~{Israelian}, M.~{Mayor}, R.~{Rebolo}, and S.~{Udry}.
\newblock {Statistical properties of exoplanets. II. Metallicity, orbital
  parameters, and space velocities}.
\newblock {\em A\&A}, 398:363--376, January 2003.

\bibitem{Santos02}
N.~C. {Santos}, M.~{Mayor}, D.~{Naef}, F.~{Pepe}, D.~{Queloz}, S.~{Udry},
  M.~{Burnet}, J.~V. {Clausen}, B.~E. {Helt}, E.~H. {Olsen}, and J.~D.
  {Pritchard}.
\newblock {The CORALIE survey for southern extra-solar planets. IX. A 1.3-day
  period brown dwarf disguised as a planet}.
\newblock {\em A\&A}, 392:215--229, September 2002.

\bibitem{Skrutskie06}
M.~F. {Skrutskie}, R.~M. {Cutri}, R.~{Stiening}, M.~D. {Weinberg},
  S.~{Schneider}, J.~M. {Carpenter}, C.~{Beichman}, R.~{Capps}, T.~{Chester},
  J.~{Elias}, J.~{Huchra}, J.~{Liebert}, C.~{Lonsdale}, D.~G. {Monet},
  S.~{Price}, P.~{Seitzer}, T.~{Jarrett}, J.~D. {Kirkpatrick}, J.~E. {Gizis},
  E.~{Howard}, T.~{Evans}, J.~{Fowler}, L.~{Fullmer}, R.~{Hurt}, R.~{Light},
  E.~L. {Kopan}, K.~A. {Marsh}, H.~L. {McCallon}, R.~{Tam}, S.~{Van Dyk}, and
  S.~{Wheelock}.
\newblock {The Two Micron All Sky Survey (2MASS)}.
\newblock {\em AJ}, 131:1163--1183, February 2006.

\bibitem{Soederhjelm99}
S.~{S{\"o}derhjelm}.
\newblock {Visual binary orbits and masses POST HIPPARCOS}.
\newblock {\em A\&A}, 341:121--140, January 1999.

\bibitem{Takeda05}
G.~{Takeda} and F.~A. {Rasio}.
\newblock {High Orbital Eccentricities of Extrasolar Planets Induced by the
  Kozai Mechanism}.
\newblock {\em ApJ}, 627:1001--1010, July 2005.

\bibitem{Thebault04}
P.~{Th{\'e}bault}, F.~{Marzari}, H.~{Scholl}, D.~{Turrini}, and M.~{Barbieri}.
\newblock {Planetary formation in the {$\gamma$} Cephei system}.
\newblock {\em A\&A}, 427:1097--1104, December 2004.

\bibitem{Udry04}
S.~{Udry}, A.~{Eggenberger}, J.-L. {Beuzit}, A.-M. {Lagrange}, M.~{Mayor}, and
  G.~{Chauvin}.
\newblock {The binarity status of stars with and without planets probed with
  VLT/NACO}.
\newblock In C.~{Allen} and C.~{Scarfe}, editors, {\em Revista Mexicana de
  Astronomia y Astrofisica Conference Series}, pages 215--216, August 2004.

\bibitem{Udry07a}
S.~{Udry}, D.~{Fischer}, and D.~{Queloz}.
\newblock {A Decade of Radial-Velocity Discoveries in the Exoplanet Domain}.
\newblock In B.~{Reipurth}, D.~{Jewitt}, and K.~{Keil}, editors, {\em
  Protostars and Planets V}, pages 685--699, 2007.

\bibitem{Udry02}
S.~{Udry}, M.~{Mayor}, D.~{Naef}, F.~{Pepe}, D.~{Queloz}, N.~C. {Santos}, and
  M.~{Burnet}.
\newblock {The CORALIE survey for southern extra-solar planets. VIII. The very
  low-mass companions of HD 141937, HD 162020, HD 168443 and HD 202206: Brown
  dwarfs or ``superplanets''?}
\newblock {\em A\&A}, 390:267--279, July 2002.

\bibitem{Udry00}
S.~{Udry}, M.~{Mayor}, D.~{Naef}, F.~{Pepe}, D.~{Queloz}, N.~C. {Santos},
  M.~{Burnet}, B.~{Confino}, and C.~{Melo}.
\newblock {The CORALIE survey for southern extra-solar planets. II. The
  short-period planetary companions to HD 75289 and HD 130322}.
\newblock {\em A\&A}, 356:590--598, April 2000.

\bibitem{Udry03}
S.~{Udry}, M.~{Mayor}, and N.~C. {Santos}.
\newblock {Statistical properties of exoplanets. I. The period distribution:
  Constraints for the migration scenario}.
\newblock {\em A\&A}, 407:369--376, August 2003.

\bibitem{Udry07b}
S.~{Udry} and N.C. {Santos}.
\newblock {Statistical Properties of Exoplanets}.
\newblock {\em ARAA}, 2007.
\newblock in press.

\bibitem{Wu03}
Y.~{Wu} and N.~{Murray}.
\newblock {Planet Migration and Binary Companions: The Case of HD 80606b}.
\newblock {\em ApJquit}, 589:605--614, May 2003.

\bibitem{Zucker94}
S.~{Zucker} and T.~{Mazeh}.
\newblock {Study of spectroscopic binaries with TODCOR. 1: A new
  two-dimensional correlation algorithm to derive the radial velocities of the
  two components}.
\newblock {\em ApJ}, 420:806--810, January 1994.

\bibitem{Zucker02}
S.~{Zucker} and T.~{Mazeh}.
\newblock {On the Mass-Period Correlation of the Extrasolar Planets}.
\newblock {\em ApJL}, 568:L113--L116, April 2002.

\bibitem{Zucker03}
S.~{Zucker}, T.~{Mazeh}, N.~C. {Santos}, S.~{Udry}, and M.~{Mayor}.
\newblock {Multi-order TODCOR: Application to observations taken with the
  CORALIE echelle spectrograph. I. The system HD 41004}.
\newblock {\em A\&A}, 404:775--781, June 2003.

\bibitem{Zucker04}
S.~{Zucker}, T.~{Mazeh}, N.~C. {Santos}, S.~{Udry}, and M.~{Mayor}.
\newblock {Multi-order TODCOR: Application to observations taken with the
  CORALIE echelle spectrograph. II. A planet in the system HD 41004}.
\newblock {\em A\&A}, 426:695--698, November 2004.

\end{thebibliography}
%

\printindex
\end{document}